\documentclass[prd,aps,showpacs,nofootinbib,superscriptaddress,amsmath,amssymb,eqsecnum,preprint,tightenlines]{revtex4}

\usepackage{graphicx}
\usepackage{bm} 
\usepackage{epsfig}
\usepackage[latin1]{inputenc}
\usepackage{float,amsmath}

\def\lsim{\mathrel{\rlap{\lower4pt\hbox{$\sim$}}
    \raise1pt\hbox{$<$}}}                
\def\gsim{\mathrel{\rlap{\lower4pt\hbox{$\sim$}}
    \raise1pt\hbox{$>$}}}                
\newcommand{\beq}{\begin{equation}}
\newcommand{\eeq}{\end{equation}}
\newcommand{\bqa}{\begin{eqnarray}}
\newcommand{\eqa}{\end{eqnarray}}

\begin{document}


\title{Constraining relativistic viscous hydrodynamical evolution}

\author{Mauricio Martinez}
\affiliation{
Helmholtz Research School and Otto Stern School\\
  Goethe - Universit\"at Frankfurt am Main\\
  Ruth-Moufang-Str.\,1,\\
  60438 Frankfurt am Main, Germany
}
\author{Michael Strickland}
\affiliation{
Physics Department, Gettysburg College\\
  Gettysburg, PA 17325 United States\\
  \vspace{5mm}
}

\begin{abstract}
{
We show that by requiring positivity of the longitudinal pressure 
it is possible to constrain the initial conditions one can use in
2nd-order viscous hydrodynamical simulations of ultrarelativistic heavy-ion 
collisions.  We demonstrate this
explicitly for 0+1 dimensional viscous hydrodynamics and discuss how
the constraint extends to higher dimensions.  Additionally, we present 
an analytic approximation to the solution of 0+1 dimensional
2nd-order viscous hydrodynamical evolution equations appropriate
to describe the evolution of matter in an ultrarelativistic heavy-ion 
collision.
}
\end{abstract}
\maketitle

\section{Introduction}
\label{sec:intro}

The main goal of experiments which perform ultrarelativistic heavy-ion collisions is to produce and 
study the properties of a deconfined plasma of quarks and gluons. This new state of matter,
the quark-gluon plasma (QGP), is expected to be formed once the temperature of nuclear matter exceeds a critical 
temperature of $T_C\sim$ 200 MeV.  Such experiments have already been underway for
nearly a decade at the Relativistic Heavy Ion Collider (RHIC) and higher-energy runs 
are planned at Large Hadron Collider (LHC).  Historically, in order to make phenomenological predictions for 
experimental observables, fluid hydrodynamics has been used to model the space-time evolution and 
non-equilibrium properties of the expanding matter. 
For the description of nuclear matter by fluid hydrodynamics to be valid the 
microscopic interaction time scale must be much shorter than the macroscopic evolution time scale. However, the 
hot and dense matter created in these experiments is rather small in transverse extent and expands very rapidly
causing the range of validity of hydrodynamics to be limited. 

After the first results of RHIC, it was somewhat of a surprise that ideal hydrodynamics could 
reproduce the hadron transverse momentum spectra 
in central and semi-peripheral collisions.  This included their anisotropy in non-central collisions which is 
measured by the elliptic flow coefficient, 
$v_2 (p_T)$.  Ideal hydrodynamical models were fairly successful in describing the dependence of $v_2$
on the hadron rest mass for transverse momenta up to about 1.5-2 GeV/c 
\cite{Huovinen:2001cy, Hirano:2002ds,Tannenbaum:2006ch, Kolb:2003dz}. This observation led to the conclusion that the QGP 
formed at RHIC could have a short thermalization time ($\tau_0 \lsim 1$ fm/c) and a low shear viscosity.  As a result 
it was posited that the matter created in the 
experiment behaves like a nearly perfect fluid starting at very early times after the collision. 
However, recent results from viscous 
hydrodynamical simulations which include all 2nd-order transport coefficients consistent with
conformal symmetry \cite{Luzum:2008cw} have 
shown that estimates of the thermalization time are rather uncertain due to poor knowledge of the proper initial 
conditions, details of plasma hadronization, subsequent hadronic cascade, 
etc.\footnote{For more about the application 
of viscous hydrodynamics to heavy-ion phenomenology we refer the reader to 
Refs.~\cite{Dusling:2007gi,Luzum:2008cw,Song:2008hj,Heinz:2009xj}.}. As a result, it now seems that thermalization times of 
up to $\tau_0\sim 2$  fm/c are 
not completely ruled out by RHIC data.  Faced with this challenge it has been recently suggested that it may be
possible to experimentally constrain $\tau_0$ by making use of high-energy electromagnetic probes such as dileptons \cite{Mauricio:2007vz, Martinez:2008di, Martinez:2008rr, Martinez:2008mc} and 
photons \cite{Schenke:2006yp, Bhattacharya:2008up, Bhattacharya:2008mv}.

As mentioned above, 
one of the key ingredients necessary to perform any numerical simulation using fluid hydrodynamics is the proper choice of
initial conditions at the initially simulated time ($\tau_0$).  These initial conditions include the initial fluid energy density 
$\epsilon$, the initial components of the fluid velocity  $u^\mu$ and the initial shear tensor $\Pi^{\mu\nu}$. 
Once the set of initial conditions is known, it is ``simple'' to 
follow the subsequent dynamics of the fluid equations in simulations. At the moment there is no first principles
calculation that 
allows one to determine the initial conditions necessary. Two different approaches are currently 
used for numerical simulations of fluids in 
heavy-ion collisions: Glauber type \cite{Kolb:2001qz} or Colored-Glass-Condensate (CGC) 
initial conditions\footnote{For a recent review on the initial conditions based on the 
CGC approach see Ref.~\cite{Lappi:2009mp} and references 
therein.}. The uncertainty in the initial conditions introduces a systematic theoretical uncertainty when, for example, 
the transport coefficient 
$\eta / s$ is extracted from experimental data \cite{Dusling:2007gi,Luzum:2008cw,Song:2008hj,Heinz:2009xj}. This is due to the fact that 
when the initial energy density profile is 
fixed using CGC-based initial conditions \cite{Hirano:2005xf, Drescher:2006pi,Lappi:2006xc}, one obtains larger 
initial spatial eccentricity and momentum anisotropy when compared with the Glauber model. Moreover, the values of the 
components of the shear tensor $\Pi^{\mu\nu}$ at $\tau_0$ are also affected by the choice of either CGC or 
Glauber initial conditions (see discusion in Sect. 4 of Ref. \cite{Baier:2006gy}).  In the case of Glauber initial
conditions the shear tensor is completely unconstrained.  In the case of CGC initial
conditions there is a prescription for calculating the initial shear; however, with CGC initial conditions the
longitudinal pressure is zero due to the assumption of exact boost invariance and the subsequent thermalization of
the system could completely change the initial shear obtained in the CGC approximation.  Therefore, in both 
cases it would seem that the initial shear is completely unconstrained.

Given these uncertainties it would be useful to have a method which can help to constrain the
allowed initial conditions used in hydrodynamical simulations.
In this work we derive general criteria which impose bounds on the initial time $\tau_0$ at which one can apply 
2nd-order viscous
hydrodynamical modeling of the matter created in ultrarelativistic heavy-ion collision.  We do this firstly 
by requiring the positivity of the effective longitudinal pressure and secondly by requiring that the shear tensor
be small compared to the isotropic pressure.  Based on these requirements we find that, for a given set of transport coefficients,
the allowed minimum value 
of $\tau_{0}$ is non-trivially related with the initial condition for the shear tensor, $\Pi^{\mu\nu} (\tau_0)\equiv 
\Pi^{\mu\nu}_0$, and the energy density $\epsilon (\tau_0)\equiv\epsilon_0$. To make this explicit we study 
$0+1$ dimensional 2nd-order viscous
hydrodynamics \cite{Muronga:2003ta, Baier:2007ix, Bhattacharyya:2008jc}, where the transport coefficients are either those of a 
weakly-coupled transport theory \cite{Arnold:2000dr, Arnold:2003zc,York:2008rr} 
or those obtained from a strongly-coupled ${\cal N}=4$ supersymmetric (SYM) plasma 
\cite{Baier:2007ix, Bhattacharyya:2008jc}.  We then show how the constraints derived from the 0+1 dimensional
case can be used to estimate where higher dimensional simulations will cease to be physical/trustworthy.
Our technique is complementary to the approach of Molnar and Huovinen \cite{Huovinen:2008te} which
uses kinetic theory to assess the applicability of hydrodynamics.  In contrast to their work, here we do not invoke any
other physics other than hydrodynamical evolution itself and merely require that it be reasonably self-consistent.

The work is organized as follows: in Sec.~\ref{sec:setup} we review the basic setup of 2nd-order viscous 
hydrodynamics formalism and its application to a $0+1$ dimensional boost invariant QGP (either in the weakly or strongly 
coupled limits). In Sec.~\ref{sec:analyticapproximation} we present an approximate analytical solution 
to the equations of motion for a $0+1$ dimensional system. In Sec.~\ref{sec:results}, we 
present our analytical and numerical results in both the strong and weak coupling limits of the 
$0+1$ dimensional QGP.  In Sec.~\ref{sec:conclusions} we present our conclusions.

\section{Basic setup}
\label{sec:setup}

In this section we briefly review the general framework of 2nd-order viscous hydrodynamics 
equations for a conformal fluid, i.e. we will consider just shear viscosity and neglect bulk 
viscosity.  We will also ignore heat conduction. The energy-momentum tensor for a relativistic fluid in 
the presence of shear viscosity is given by\footnote{The notation we use along the text is summarized in the Appendix 
\ref{app:notations}.}:
\begin{equation}
\label{tensor}
 T^{\mu \nu}=\epsilon\, u^\mu\, u^\nu-p \Delta^{\mu \nu}+\Pi^{\mu \nu}, 
\end{equation}
where $\epsilon$ and $p$ are the fluid energy density and pressure, $u^\mu$ is the normalized fluid four-velocity 
($u^\mu u_\mu =1$) and $\Pi^{\mu \nu}$ is the shear tensor which has two important properties: 
(1) $\Pi^\mu_\mu =0$ and (2) $u_\mu\Pi^{\mu \nu}=0$. Requiring 
conservation of energy and momentum, $D_\mu T^{\mu \alpha}=0$, gives the space-time evolution equations for 
the fluid velocity and the energy density:
\begin{eqnarray}
\label{eqsvel+ene}
(\epsilon+p)D u^\mu&=&\nabla^\mu p-
\Delta^\mu_\alpha D_\beta \Pi^{\alpha \beta}\, ,
\nonumber\\
D \epsilon &=& - (\epsilon+p) \nabla_\mu u^\mu+\frac{1}{2}\Pi^{\mu \nu}
\nabla_{\langle \nu} u_{\mu\rangle}\, ,
\end{eqnarray}
where $D_\mu$ is the geometric covariant derivative,
$D\equiv u^\alpha D_\alpha$ is the comoving time derivative in the fluid rest frame and 
$\nabla^\mu\equiv \Delta^{\mu \alpha} D_\alpha$ is the spatial derivative in the fluid rest frame. 
The brackets $\langle\ \rangle$ construct terms which are 
symmetric, traceless, and orthogonal to the fluid velocity (see Appendix 
\ref{app:notations} for its definition). 

To obtain a complete solvable system of equations viscous hydrodynamics requires an additional equation of motion for the shear 
tensor. This is accomplished by expanding the equations of motion to 
second order in gradients.  It has been found that at zero-chemical potential in a conformal fluid in any curved space-time, 
the shear tensor satisfies \cite{Baier:2007ix, Bhattacharyya:2008jc}:
\begin{eqnarray}
 \Pi^{\mu\nu} &=& \eta \nabla^{\langle \mu} u^{\nu\rangle}
- \tau_\pi \left[ \Delta^\mu_\alpha \Delta^\nu_\beta D\Pi^{\alpha\beta} 
 + \frac 4{3} \Pi^{\mu\nu}
    (\nabla_\alpha u^\alpha) \right] \nonumber\\
  &&\quad 
  + \frac{\kappa}{2}\left[R^{<\mu\nu>}+2 u_\alpha R^{\alpha<\mu\nu>\beta} 
      u_\beta\right]\nonumber\\
  && -\frac{\lambda_1}{2\eta^2} {\Pi^{<\mu}}_\lambda \Pi^{\nu>\lambda}
  +\frac{\lambda_2}{2\eta} {\Pi^{<\mu}}_\lambda \omega^{\nu>\lambda}
  - \frac{\lambda_3}{2} {\omega^{<\mu}}_\lambda \omega^{\nu>\lambda}\, ,
\label{pieq}
\end{eqnarray}
where $\omega_{\mu \nu}=-\nabla_{[\mu} u_{\nu]}$ is a symmetric operator that represents the 
fluid vorticity and $R^{\alpha \mu \nu \beta}$ and $R^{\mu \nu}$
are the Riemann and Ricci tensors, respectively.
The coefficients $\tau_\pi,\kappa,\lambda_1,\lambda_2$ and $\lambda_3$
are the transport coefficients required by conformal symmetry.

\subsection{0+1 Dimensional Conformal 2nd-Order Viscous Hydrodynamics}
\label{subsec:hydroeqs}

Let us consider a system expanding in a boost invariant manner along the longitudinal (beamline)
direction with a uniform energy density along the transverse plane. For this simplest heavy-ion collision model, 
it is enough to consider expansion in a flat space. Also for this 
simple model, there is no fluid vorticity, and the energy density, the shear viscous tensor and the fluid velocity 
only depend on proper time $\tau$. For this 0+1 dimensional model the
 2nd-order viscous hydrodynamic equations (Eqs.~(\ref{eqsvel+ene}) and (\ref{pieq})) are rather simple 
in the conformal limit. In terms of proper time, $\tau =
\sqrt{t^2 -z^2}$, and space-time rapidity, $\zeta = {\rm arctanh}(z/t)$, these are given by 
\cite{Muronga:2003ta,Baier:2007ix}:
\bqa
\partial_\tau \epsilon&=&-\frac{\epsilon+p}{\tau}+\frac{\Pi}{\tau} \, ,
\label{0+1eqe}\\
\partial_\tau \Pi &=& -\frac{\Pi}{\tau_\pi}
+\frac{4 \eta}{3\, \tau_\pi \tau}-\frac{4}{3\, \tau} \Pi
-\frac{\lambda_1}{2\,\tau_\pi\,\eta^2} \left(\Pi\right)^2 \, ,
\label{0+1eqp}
\eqa
where $\epsilon$ is the fluid energy density, $p$ is the fluid pressure, 
$\Pi \equiv \Pi^\zeta_\zeta$ is the $\zeta\zeta$ component 
of the fluid shear tensor, $\eta$ is the fluid shear viscosity, $\tau_\pi$ is the shear 
relaxation time, and $\lambda_1$ is a coefficient which arises in complete 2nd-order 
viscous hydrodynamical equations either in the strong 
\cite{Baier:2007ix,Bhattacharyya:2008jc} or weakly coupled limit 
\cite{Arnold:2000dr, Arnold:2003zc,Muronga:2003ta,York:2008rr,Betz:2008me}.
The Navier-Stokes limit is recovered upon taking $\tau_\pi\rightarrow 0$ and
$ \lambda_1 \rightarrow 0$ in which case one obtains
$\Pi_\text{Navier-Stokes} = 4 \eta/(3 \tau)$.

These coupled differential equations are completed by a specification of the
equation of state which relates the energy density and the pressure through 
$p = p(\epsilon)$ and initial conditions.  For 0+1 dimensional dynamics one
must specify the energy 
density and $\Pi$ at the initial time, $\epsilon_0 \equiv \epsilon(\tau_0)$
and $\Pi_0 \equiv \Pi(\tau_0)$, where $\tau_0$ is the proper-time at which one 
begins to solve the differential equations.

\subsection{Specification of equation of state and dimensionless variables}
\label{subsec:eosscaling}

In the following analysis we will assume an ideal equation of state, in which case
we have
\beq 
\label{eqstate}
p = \frac{N_{\rm dof}\, \pi^2}{90} T^4\, ,
\eeq
where for quantum chromodynamics with $N_c$ colors and $N_f$ quark flavors,
$N_{\rm dof} = 2 (N_c^2-1) + 7 N_c N_f/2$ which for $N_c=3$ and $N_f=2$ 
is $N_{\rm dof} = 37$.  The general method used below, however, can easily be extended
to a more realistic equation of state. 

In the conformal limit the trace of the four-dimensional stress tensor vanishes requiring
$\epsilon = 3 p$ which, using Eq.~(\ref{eqstate}), allows us to write compactly
\beq
\epsilon = (T/\gamma)^4, \hspace{0.5cm} \text{with}\hspace{0.5cm}\gamma \equiv \left( \frac{30}{\pi^2 N_{\rm dof}} 
\right)^{1/4} \,.
\label{energyideal}
\eeq
Likewise we can simplify the expression for the entropy density, $s$, using the thermodynamic relation 
$T s = \epsilon + p$ to obtain $s = 4 \epsilon / 3 T$ or equivalently
\beq
s = \frac{4}{3 \gamma}\, \epsilon^{3/4} \, .
\label{entropyideal}
\eeq

\begin{table}[t]
\begin{tabular}{ | c || c | c | c |}
\hline
{\bf $\;$ Transport coefficient $\;$} & 
{\bf $\;$ Weakly-coupled QCD $\;$ } & 
{\bf $\;$ Strongly-coupled ${\cal N}=4$ SYM $\;$} \\ \hline
$\bar{\eta}\equiv \eta/s$ & $\thicksim1/(g^4 \log g)$ & $1/(4 \pi)$\\ \hline
$\tau_\pi$ & $6 \bar{\eta}/T$ & $\bigl(2-\log 2\bigr)/(2\pi T)$ \\ \hline
$\lambda_1$ & $(4.1 \rightarrow 5.2)\,\bar{\eta}^2 s/T$ &  2 $\bar{\eta}^2 s/T$\\ \hline
\end{tabular}
\caption{Typical values of the transport coefficients for a weakly-coupled QGP 
\cite{York:2008rr,Arnold:2000dr,Arnold:2003zc} and 
a strongly coupled ${\cal N}=4$ SYM plasma \cite{Baier:2007ix,Bhattacharyya:2008jc}.}
\label{transcoeff}
\end{table}

When solving Eqs.~(\ref{0+1eqe}) and (\ref{0+1eqp}) it is important to recognize that 
the transport coefficients depend on the temperature of the plasma and hence on 
proper-time. We summarize in Table \ref{transcoeff} the values of the transport coefficients in the 
strong and weak coupling limits.
We point out that these are not universal relations as explained below in Secs.~\ref{strongcouplinglimit} and 
\ref{weakcouplinglimit}. The reader should note that in 
either the strong or weak coupling limit the coefficients $\tau_\pi$ and $\lambda_1$ are proportional
to $\tau_\pi\propto T^{-1}$ and $
\lambda_1\propto \bar\eta^2 s / T$. This suggests that we can parametrize both coefficients as:
\begin{subequations}
 \label{parametrization}
\begin{align}
 \tau_\pi & =  \frac{c_\pi}{T} \, ,\\
 \lambda_1 & =  c_{\lambda_1} \bar{\eta}^2 \biggl(\frac{s}{T}\biggr)\, ,
\end{align}
\end{subequations}
where we have introduced a dimensionless version of the shear viscosity 
\beq
\bar{\eta}\equiv \eta / s \, .
\eeq 
In our analysis we assume that $\bar\eta$ is 
independent of time.\footnote{Including a temperature-dependent shear viscosity does not change our observations
fundamentally; however, there will be quantitative effects which will be elaborated upon in a
forthcoming publication.} The dimensionless numbers $\bar\eta$, $c_\pi$ and $c_{\lambda_1}$ carry all of the
information about the particular coupling limit we are 
considering. 

Using the ideal gas equation of state [Eqs. (\ref{energyideal}) and (\ref{entropyideal})], the parametrization 
(\ref{parametrization}) of $\tau_\pi$ and $\lambda_1$ can be rewritten in terms of the energy density $\epsilon$:

\begin{subequations}
\label{parametrization2}
 \begin{align}
 \tau_\pi & =  \frac{c_\pi}{\gamma\, \epsilon^{1/4}} \, ,\\  
 \lambda_1 & =  \frac{4}{3\gamma^2}\, c_{\lambda_1}\, \bar{\eta}^2 \, \epsilon^{1/2} \, .
 \end{align}
\end{subequations}

To remove the dimensionful scales and rewrite the fluid equations in
a more explicit form we define the following dimensionless variables:
\begin{subequations}
\label{variabledefs}
 \begin{align}
\bar\epsilon &\equiv \epsilon/\epsilon_0  \, , \\
\overline\Pi &\equiv \Pi/\epsilon_0    \, , \\
\bar\tau &\equiv \tau/\tau_0 \, ,
\end{align}
\end{subequations}
where $\tau_0$ is the proper-time at which the hydrodynamic evolution
equations start to be integrated and $\epsilon_0$ is the energy density at $\tau_0$.

After replacing the dimensionless variables (\ref{variabledefs})
 in the parametrization (\ref{parametrization2}) and Eqs.~(\ref{0+1eqe}) and (\ref{0+1eqp}), we rewrite the fluid equations:
\begin{subequations}
\begin{eqnarray}
&& \bar\tau\, \partial_{\bar\tau} \bar\epsilon + \frac{4}{3}\, \bar\epsilon - \overline\Pi = 0  \, ,  \label{diffeqa} \\
&& \overline\Pi + \frac{c_\pi}{\gamma\, k\, \bar\epsilon^{1/4}} \left[ \partial_{\bar\tau} \overline\Pi 
      + \frac{4}{3} \frac{\overline\Pi}{\bar\tau} \right] 
      - \frac{16 \,\bar\eta}{9\, \gamma\, k} \frac{\bar\epsilon^{3/4}}{\bar\tau}
      + \frac{3\, c_{\lambda_1}}{8} \frac{{\overline\Pi}^2}{\bar\epsilon} = 0 \, , \label{diffeqb}
\end{eqnarray}
\label{diffeqs}
\end{subequations}
where $k \equiv \tau_0 \epsilon_0^{1/4}$.
Note that in terms of (\ref{variabledefs}) the boundary conditions are specified
at $\bar\tau=1$ where $\bar\epsilon=1$ and $\overline\Pi(\bar\tau=1) 
= \overline\Pi_0$ which is a free parameter.  When the hydrodynamical 
equations are written in the form given
above [Eq.~(\ref{diffeqs})] all information
about the initial proper-time and energy density is encoded in the parameter $k$ and 
all information about the equation of state is encoded in the parameter $\gamma$.

\subsection{Strong coupling limit}
\label{strongcouplinglimit}

Motivated and guided by the AdS/CFT correspondence Baier et.~al \cite{Baier:2007ix} and the Tata group 
\cite{Bhattacharyya:2008jc} have recently shown that new transport coefficients arise in a complete theory 
of second order relativistic 
viscous hydrodynamics. They also estimate their values at infinite t'Hooft coupling for ${\cal N}=4$ SYM theory 
at finite temperature.  Different calculations for a finite t'Hooft coupling within the same theory have been carried out 
\cite{Buchel:2004di,Benincasa:2005qc,Buchel:2008ac, Myers:2008yi,Paulos:2008tn,Buchel:2008bz}. A remarkable aspect 
is that, while at first the strong t'Hooft coupling limit of the transport coefficients
was expected to be universal \cite{Policastro:2001yc,Son:2007vk}, there is now evidence that 
these coefficients are not universal \cite{Brigante:2007nu,Brigante:2008gz,Kats:2007mq,%
Natsuume:2007ty,Buchel:2008vz}.  Faced with this complication one is forced to make a choice as to which
dual theory to consider.  Here we 
will consider the values obtained in ${\cal N}=4$ SYM at infinite t'Hooft coupling as used in 
\cite{Baier:2007ix,Bhattacharyya:2008jc} as our typical strong coupling values. One can expect that these coefficients change 
in strongly-coupled QCD compared to ${\cal N}=4$ SYM theory at infinite t'Hooft limit. Nevertheless, we take these 
values over from strongly-coupled ${\cal N}=4$ SYM in order to get a feeling for what to expect in this regime.

Expressed in terms of the dimensionless transport coefficients 
defined above, typical values of the strongly coupled transport coefficients are
\begin{equation}
\label{stronglimitvalues}
\begin{aligned}
\bar\eta &= \frac{1}{4 \pi} \, , \\
c_\pi &= \frac{2 - \log 2}{2 \pi} \, ,\\
c_{\lambda_1} &= 2 \, .
\end{aligned}
\end{equation}

\subsection{Weak coupling limit}
\label{weakcouplinglimit}

Contrary to the case of ${\cal N}=4$ SYM at infinite coupling, in the case of QCD,
where there is a running coupling and inherent 
scale dependence, the various transport coefficients are not fixed numbers but instead depend
on the renormalization scale.  In this limit the transport coefficients necessary have been calculated 
completely to leading order  \cite{Arnold:2000dr, Arnold:2003zc,York:2008rr}. Higher order corrections 
to some transport coefficients from 
finite-temperature perturbation theory show poor convergence \cite{CaronHuot:2008uh,CaronHuot:2007gq}
which is similar to the case for the thermodynamical potential; however, resummation techniques can dramatically
extend the range of convergence of finite-temperature
perturbation theory in the case of static quantities and can, in the future, also be applied to dynamical quantities\footnote{See 
Ref.~\cite{Andersen:2004fp} and references therein.}.  
Until such resummation schemes are carried out for dynamical quantities, the 
values of the leading-order weak-coupling transport coefficients in Table \ref{transcoeff} 
can only be considered as rough guides to the values expected phenomenologically.  Using this rough guide
the value of $\bar\eta$ from finite-temperature QCD calculations \cite{Arnold:2000dr, Arnold:2003zc} is 
$\eta / s\thicksim 0.5 \rightarrow 1$ at realistic couplings ($g\sim2\rightarrow3$).
In this work we will assume a typical value of $\bar\eta =10/(4 \pi)$ in the weakly-coupled limit
in order to compare with the results obtained in the 
strong coupling limit. In our analysis for the weak coupling limit, we will use

\begin{equation}
\label{weaklimitvalues}
\begin{aligned}
\bar\eta &= \frac{10}{4 \pi} \, , \\
c_\pi &= 6 \bar\eta  \, , \\
c_{\lambda_1} &= \frac{9}{2} \, .
\end{aligned}
\end{equation}

\subsection{Momentum space anisotropy}
\label{sec:momentumanisotropybounds}

We introduce the dimensionless parameter, $\Delta$, which measures 
the degree of momentum-space isotropy of the fluid as follows
\beq
\Delta \equiv \frac{p_T}{p_L} - 1 \, ,
\eeq
where $p_T = (T^{xx} + T^{yy})/2$ and $p_L = T^{zz} = - T^\zeta_\zeta$ are the effective
transverse and longitudinal pressures, respectively.  If $\Delta=0$, the 
system is locally isotropic.  If $-1 < \Delta < 0$ the system has a local prolate 
anisotropy in momentum space and if $\Delta >0$ the system has a local 
oblate anisotropy in momentum space.  In appendix \ref{app:xideltarelation}
we derive the relation between the $\Delta$ parameter defined above and
the $\xi$ parameter introduced in Ref.~\cite{Romatschke:2003ms} to quantify 
the degree of local plasma isotropy.  For small values of $\Delta$ the 
relation is $\Delta = 4\xi/5 + {\cal O}(\xi^2)$.

In the 0+1 dimensional model of viscous hydrodynamics one can express the 
effective transverse pressure as $p_T = p + \Pi/2$ and the 
effective longitudinal pressure as $p_L = p - \Pi$.  In the case of an ideal
equation of state, rewriting (\ref{variabledefs}) in terms of
our dimensionless variables gives %
\beq
\Delta = \frac{9}{2} \left(\frac{\overline\Pi}{\bar\epsilon - 3 \overline\Pi}\right) \, .
\eeq
At the initial time $\bar\tau=1$, $\Delta_0 \equiv \Delta(\bar\tau=1)$ is given by
\beq
\Delta_0 = \frac{9}{2} \left(\frac{\overline\Pi_0}{1 - 3 \overline\Pi_0}\right) \, .
\eeq
In the limit $\overline\Pi \rightarrow -2\bar\epsilon/3$ we have
$\Delta \rightarrow -1$ and in the limit
$\overline\Pi \rightarrow \bar\epsilon/3$ we have $\Delta \rightarrow \infty$.

Positivity of the longitudinal pressure requires $\Delta \neq \infty$ at any time
during the evolution of the plasma.  Note that requiring positivity is a {\em
weak constraint} on the magnitude of $\Delta$ since the formal justification for
applying viscous hydrodynamical approximations is the neglect of large
gradients and higher-order nonlinear terms.  This requires that ${\overline\Pi}$
be small compared to the pressure, $p$, i.e. $|{\overline\Pi}| \ll \bar{p}$.  This
can be turned into a quantitative statement by requiring that  $-\alpha \, \bar{p} < {\overline\Pi} 
< \alpha \, \bar{p}$, where $\alpha$ is a positive phenomenological constant 
which is less than or equal to 1, i.e. $0 \leq \alpha \leq 1$.  The limit 
$\alpha \rightarrow 1$ gives the weak constraint of $-3/4 \leq \Delta < \infty$ and for 
general $\alpha$ requires $\Delta_-  \leq \Delta \leq \Delta_+$ where 
\beq
\Delta_\pm \equiv \pm\frac{3}{2} \left(\frac{\alpha}{1 \mp \alpha}\right) \, .
\eeq
For example, requiring $\alpha = 1/3$ we would find the constraint
$-3/8 \leq \Delta_\alpha \leq 3/4$.

\section{Approximate Analytic Solution of 0+1 Conformal Hydrodynamics}
\label{sec:analyticapproximation}

In this section we present an approximate analytic solution to the 0+1 dimensional
conformal 2nd-order
hydrodynamical evolution equations.  The approximation used will be to
first exactly integrate the differential equation for the energy density (\ref{diffeqa}), 
thereby expressing the energy density
as an integral of the shear.  We then insert this integral relation into the equation
of motion for shear itself (\ref{diffeqb}) and expand in $\bar{\eta}$.  Explicitly, the solution 
obtained  from the first step is
\beq
\bar{\epsilon}(\bar\tau) = \bar\tau^{-4/3} \left[ 1+
                         \int_1^{\bar\tau} \, d{\bar\tau^\prime} 
                          (\tau^\prime)^{1/3} {\overline\Pi}(\bar\tau^\prime)  \right]
                          \, .
\label{analyticesolution}
\eeq
We then solve the second differential equation for $\overline\Pi$ approximately 
by dropping the second term in Eq.~(\ref{analyticesolution}) and inserting this into the 
second equation of (\ref{diffeqs}) to obtain
\beq
27 \, c_{\lambda_1} \, \gamma \, k \, \bar\tau^{10/3} \, \overline\Pi^2
+ 72 \, c_\pi \, \bar\tau^{7/3} \, \partial_{\bar\tau}\overline\Pi
+ ( 72 \, \gamma  \, k  \, \bar\tau^2 + 96 \, c_\pi \, \bar\tau^{4/3}) \, \overline\Pi  = 128 \, \bar\eta \, .
\label{pilodiffeq}
\eeq
This differential equation has a solution of the form
\bqa
\overline\Pi = 
&& \left(\frac{4}{3 c_{\lambda_1} \bar\tau^{4/3}}\right) \nonumber \\
&& \times \, \frac{
  {\cal C} \left[ 2 \; {}_1F_1\left( { 1-b \atop 2  } \Big| - a \, \bar\tau^{2/3} \right) 
    + a\, (b-1) \, \bar\tau^{2/3} \; {}_1F_1\left( { 2-b \atop 3  } \Big| - a \, \bar\tau^{2/3} \right) \right] 
    + 2 \, G_{1,2}^{2,0}\left( a \, \bar\tau^{2/3} \Big| {b \atop 0,0} \right)
 }{
  a \, {\cal C} \, \bar\tau^{2/3} \; {}_1F_1\left( { 1-b \atop 2  } \Big| - a \, \bar\tau^{2/3} \right)
  - G_{1,2}^{2,0}\left( a \, \bar\tau^{2/3} \Big| {b+1 \atop 0,1} \right)
 } \, , \nonumber \\
 \label{analyticpisolution}
\eqa
where ${}_1F_1$ is a confluent hypergeometric function, $G$ is the Meijer G function,
$a = 3 \gamma k / (2 c_\pi)$, $b = c_{\lambda_1} \bar\eta  / c_\pi$, and ${\cal C}$ 
is an integration constant which is fixed by the initial condition for $\overline\Pi$ at $\bar\tau=1$.
Requiring $\overline\Pi(\bar\tau=1) = \overline\Pi_0$ fixes ${\cal C}$ to be
\beq
{\cal C} = 
\frac{
  8 \, G_{1,2}^{2,0}\left( a \Big| {b \atop 0,0} \right)
  + 3\, c_{\lambda_1} \, \overline\Pi_0 \,
     G_{1,2}^{2,0}\left( a \Big| {b+1 \atop 0,1} \right)
 }{
  \left[ 3\, a \, c_{\lambda_1} \, \overline\Pi_0 - 8 \right] 
  {}_1F_1\left( { 1-b \atop 2  } \Big| - a \right)
  - 4 \, a \, (b-1) \,
   {}_1F_1\left( { 2-b \atop 3  } \Big| - a \right)
 } \, .
 \eeq
 To obtain the proper-time evolution of the energy density one must integrate (\ref{analyticesolution})
 using (\ref{analyticpisolution}).  This is possible to do analytically but the answer is rather unwieldy and
 hence not very useful to list explicitly.  Below we will use this approximate analytic solution as a 
 cross check for our numerics.  In the limit $\bar\eta \rightarrow 0$ this solution becomes an 
 increasingly better approximation and hence represents the leading correction to ideal hydrodynamical 
 evolution in that limit.
  
 Note that in the limit $c_{\lambda_1} \rightarrow 0$ and $c_\pi \rightarrow 0$
 the differential equation above (\ref{pilodiffeq}) reduces to an algebraic equation
 \beq
 \overline\Pi_\text{Ideal Navier-Stokes} = \frac{16 \bar\eta}{9 \gamma k \bar\tau^2} \, ,
 \eeq
 which, when converted back to dimensionful variables, corresponds to the
 Navier-Stokes solution under the assumption that $\bar\epsilon = \bar\tau^{-4/3}$.  
 Finally we note that in the large time limit Eq.~(\ref{analyticpisolution}) simplifies
 to
 \beq
 \lim_{\bar\tau \rightarrow \infty} \overline\Pi = \overline\Pi_\text{Ideal Navier-Stokes} 
 	+ {\cal O}\left(e^{- a \bar\tau^{-2/3}}\right) \, .
 \eeq
 
\section{Results}
\label{sec:results}

In this section we present our results of numerical integration of Eq.~(\ref{diffeqs}) 
and present consistency checks obtained by comparing these results with the approximate
analytic solution presented in the previous section.

\subsection{Time Evolution of $\Delta$}

Below we present numerical results for the time evolution of the plasma anisotropy parameter
$\Delta$.  For purpose of illustration we will hold the initial temperature fixed at $T = $ 350 MeV
and vary the starting time $\tau_0$.  This will allow us to probe different values of 
$k = \tau_0 \epsilon_0^{1/4} = \tau_0 T_0/\gamma$ in a transparent manner.  
Note that, by doing this,
each curve corresponds to a different initial entropy density; however, this is irrelevant for the
immediate discussion since we are not concerned with phenomenological
consequences, only with the general mathematical properties of the system of differential
equations as one varies the fundamental parameters.  In Secs.~\ref{sec:criticalline} and 
\ref{sec:convergencecriteria} we will present the general results as a function of the dimensionless 
parameter $k$.

\subsubsection{Strong Coupling}

\begin{figure*}[t]
\begin{center}
\includegraphics[width=10cm]{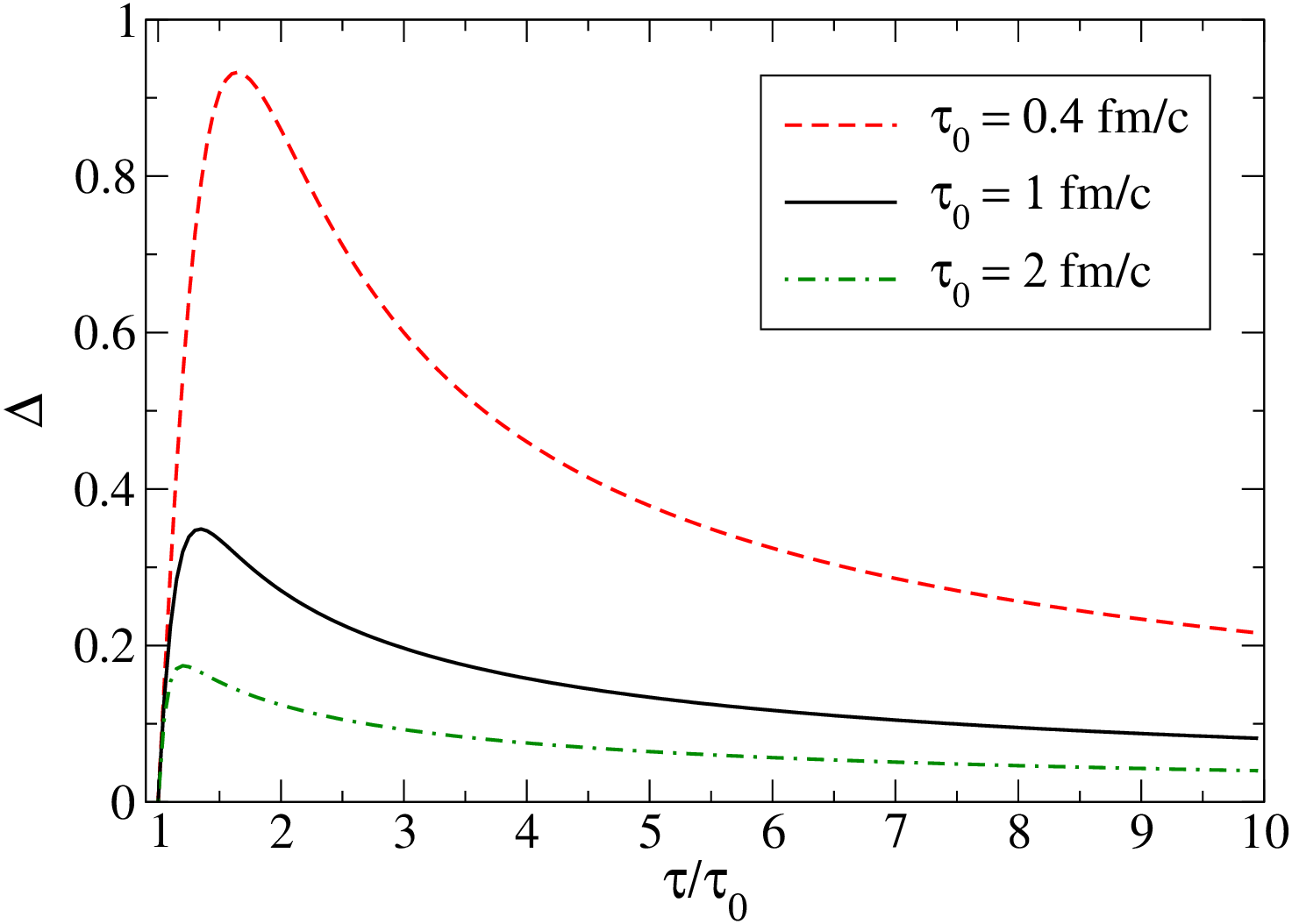}
\end{center}
\vspace{-6mm}
\caption{
Result for the proper-time evolution of $\Delta$ obtained by numerical integration of Eq.~(\ref{diffeqs}).  Long-dashed, solid, 
and short-dashed lines correspond $\tau_0 = \{0.4, 1, 2\}$ fm/c, respectively.  Transport coefficients were the typical strong coupling values given in Eq.~(\ref{stronglimitvalues}).  The initial 
temperature, $T_0$, is held fixed at $T_0 = 350$ MeV and it is assumed that $\Delta_0=0$ for this example.
}
\label{fig:strongcouplingvaryingtau0}
\end{figure*}

In Fig.~\ref{fig:strongcouplingvaryingtau0} we show our result for the proper-time 
evolution of the pressure anisotropy parameter, $\Delta$, obtained by numerical 
integration of Eq.~(\ref{diffeqs}).  The transport coefficients in this case are the 
typical strong coupling values given in Eq.~(\ref{stronglimitvalues}).
For purpose of illustration we have chosen the initial temperature, $T_0$, to be held 
fixed at $T_0 = 350$ MeV and assumed that the initial
pressure anisotropy, $\Delta_0$, vanishes, i.e. $\Delta_0=0$.

As can be seen from this figure, when the initial value of the pressure anisotropy is
taken to be zero it does not remain so.  A finite oblate pressure anisotropy is rapidly 
established due to the intrinsic longitudinal expansion of the fluid.  Depending on
the initial time at which the hydrodynamic evolution is initialized, $\Delta$ peaks in the 
range $0.2 \lsim \Delta \lsim 1$.

\subsubsection{Comparison with analytic approximation}

\begin{figure*}[t]
\begin{center}
\includegraphics[width=10cm]{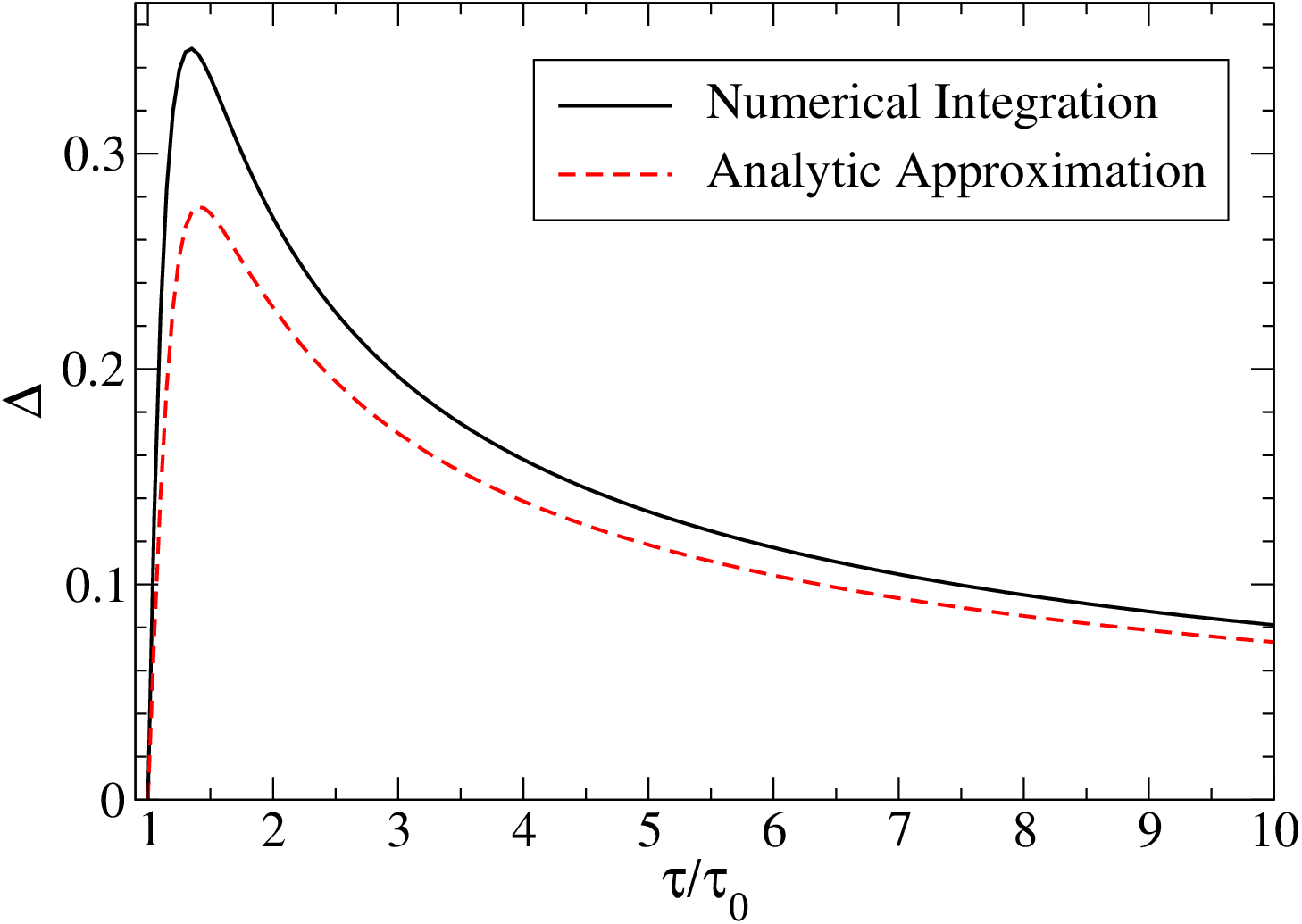}
\end{center}
\vspace{-6mm}
\caption{
Comparison of result for $\Delta$ as a function of proper time using numerical integration of Eq.~(\ref{diffeqs}) and
the approximate analytic solution given via Eqs.~(\ref{analyticpisolution}) and (\ref{analyticesolution}).  Transport 
coefficients in this case are the  typical strong coupling values given in 
Eq.~(\ref{stronglimitvalues}).  The initial temperature, $T_0$, is taken to be $T_0 = 350$ MeV, the initial time,
$\tau_0$, is taken to be $\tau_0 = 1$ fm/c and it is assumed that $\Delta_0=0$ for this example.
}
\label{fig:strongcouplingcompare}
\end{figure*}

As a cross check of our numerical method, in Fig.~\ref{fig:strongcouplingcompare} we compare
the result for $\Delta$ obtained via direct numerical integration of Eq.~(\ref{diffeqs}) and
the approximate analytic solution given via Eqs.~(\ref{analyticpisolution}) and (\ref{analyticesolution}).
As can be seen from the figure the analytic solution provides a reasonable approximation to the 
true time-evolution of the plasma anisotropy.  The parameter $\Delta$ is a particularly sensitive
quantity to compare.  If one compares the analytic and numerical solutions for the energy density,
for example, in the strongly-coupled case there is at most a 1\% deviation between the analytic
approximation and our exact numerical integration during the entire 10 fm/c of simulation time.
Of course, for larger viscosity the analytic approximation becomes more suspect but for the 
weakly-coupled case we find that there is at most a 8\% deviation between the energy densities
obtained using our analytic approximation and the exact numerical result.  In the limit that
$\bar\eta$ goes to zero, the analytic treatment and our numerical integration agree to arbitrarily
better precision.  Based on the 
agreement between the two approaches we are confident in our numerical integration of the
coupled differential equations.

\subsubsection{Weak Coupling}

\begin{figure*}[t]
\begin{center}
\includegraphics[width=10cm]{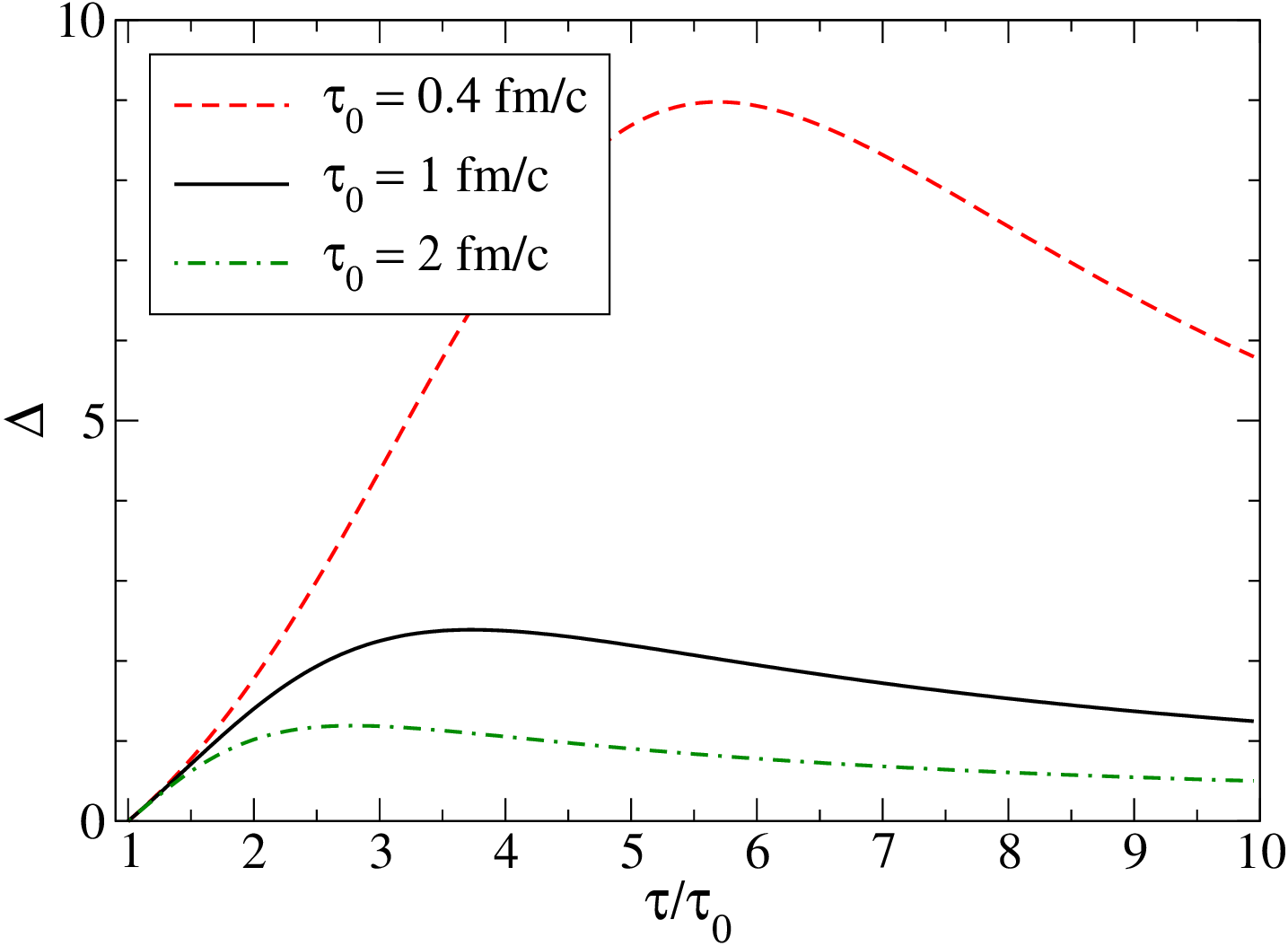}
\end{center}
\vspace{-6mm}
\caption{
Result for the proper-time evolution of $\Delta$ obtained by numerical integration of Eq.~(\ref{diffeqs}).  Long-dashed, solid, 
and short-dashed lines correspond $\tau_0 = \{0.4, 1, 2\}$ fm/c, respectively.  Transport coefficients in this case are the 
typical weak coupling values given in Eq.~(\ref{weaklimitvalues}).  The initial 
temperature, $T_0$, is held fixed at $T_0 = 350$ MeV and it is assumed that $\Delta_0=0$ for this example.
\vspace{5mm}
}
\label{fig:weakcouplingvaryingtau0}
\end{figure*}


In Fig.~\ref{fig:weakcouplingvaryingtau0} we show our result for the proper-time 
evolution of the pressure anisotropy parameter, $\Delta$, obtained by numerical 
integration of Eq.~(\ref{diffeqs}).  The transport coefficients in this case are the 
typical weak coupling values given in Eq.~(\ref{weaklimitvalues}).
For purpose of illustration we have chosen the initial 
temperature, $T_0$, to be held fixed at $T_0 = 350$ MeV and assumed that the initial
pressure anisotropy, $\Delta_0$, vanishes, i.e. $\Delta_0=0$.

As can be seen from this figure, as in the strongly coupled case, a finite oblate 
pressure anisotropy is rapidly established due to the intrinsic longitudinal expansion of the fluid.
In the case of weak coupling transport coefficients a larger pressure anisotropy develops.  
Depending on the initial time at which the hydrodynamic evolution is initialized, 
$\Delta$ peaks in the range $1 \lsim \Delta \lsim 9$.  

As can be seen from the $\tau_0 = 0.4$ fm/c result, if the initial simulation time is assumed
to be small, then very large pressure anisotropies can develop.  In that case, in dimensionful units,
the peak of the $\Delta$ evolution occurs at a time of $\tau \sim  2.3$ fm/c.  Such large 
pressure anisotropies would cast doubt on the
applicability of the 2nd-order conformal viscous hydrodynamical equations, since nonconformal
2nd-order terms and higher-order
non-linear terms corresponding to 3rd- or higher-order expansions could become 
important.\footnote{See Ref.~\cite{Betz:2008me} for an example of 2nd-order terms which
can appear when conformality is broken.}
If, in the weakly coupled case, the initial simulation time $\tau_0$ is taken to be 0.2 fm/c
one would find that $\Delta$ would become infinite during the
simulation.  This divergence is due to the fact that the longitudinal pressure goes to zero
and then becomes 
negative during some period of the time evolution.

\subsection{Negativity of Longitudinal Pressure}

\begin{figure*}[t]
\begin{center}
\includegraphics[width=10cm]{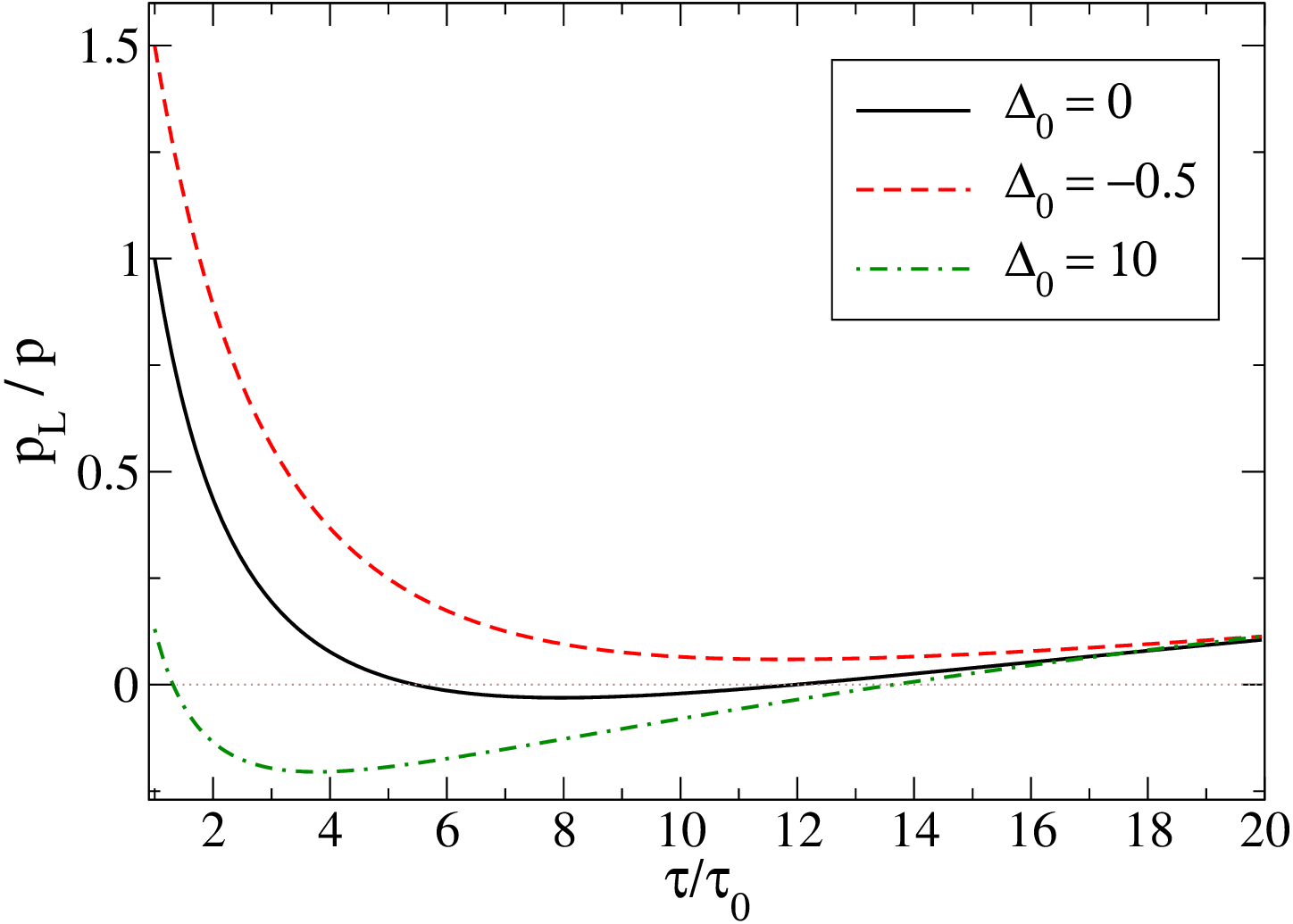}
\end{center}
\vspace{-6mm}
\caption{
Result for the proper-time evolution of the ratio of the longitudinal pressure over the pressure, $p_L/p$, 
obtained by numerical integration of Eq.~(\ref{diffeqs}).  Solid, long-dashed, and short-dashed lines correspond 
$\Delta_0 = \{0, -0.5, 10\}$, respectively.  Transport coefficients in this case are the typical 
weak coupling values given in Eq.~(\ref{weaklimitvalues}).  The initial temperature, $T_0$, is held fixed at 
$T_0 = 350$ MeV and it is assumed that $\tau_0=0.2$ fm/c for this example.  The dotted grey line indicates 
$p_L=0$ in order to more easily identify the point in time where the longitudinal pressure becomes negative.
}
\label{fig:weakcouplingvaryingxi0}
\end{figure*}

In order to explicitly demonstrate  the possibility that $\Delta$ diverges, in Fig.~\ref{fig:weakcouplingvaryingxi0}
we have plotted the evolution the longitudinal pressure 
over the isotropic pressure ($p = \epsilon/3$), $p_L/p$,  obtained by numerical integration of Eq.~(\ref{diffeqs}) 
for different assumed initial pressure anisotropies.  The transport coefficients in this case are 
the  typical weak coupling values given in Eq.~(\ref{weaklimitvalues}).  
The initial temperature, $T_0$, is held fixed at $T_0 = 350$ MeV and it is assumed that $\tau_0=0.2$ fm/c
for this example. 

As this figure shows, if the initial simulation time is too early, the longitudinal pressure
of the system can become negative.  The exact point in time at which it becomes negative depends
on the assumed initial pressure anisotropy.  As the initial pressure anisotropy becomes more prolate, 
the time over which the longitudinal pressure remains positive is increased.  For initially extremely prolate
distributions the longitudinal pressure can remain positive during the entire simulation time.  
In the opposite limit of
extremely oblate distributions, the longitudinal pressure can become negative very rapidly and remain
so throughout the entire lifetime of the plasma.  We note that in the Navier-Stokes limit  
the initial shear would be $\left(\overline\Pi_0\right)_\text{Navier Stokes}  = 16 \bar\eta/(9 \tau_0 T_0)$
which, using the initial conditions indicated in 
Fig.~\ref{fig:weakcouplingvaryingxi0}, 
gives $p_{L,0}/p = -11.1$.  This means that if one were to use Navier-Stokes initial conditions 
the system would start with an extremely large
negative longitudinal pressure.  Using $\tau_0 = 1$ fm/c and $T_0 = $350 MeV improves
the situation somewhat; however, even in that case the initial Navier-Stokes longitudinal
pressure remains negative with $p_{L,0}/p = -1.4$.

What does a negative longitudinal pressure indicate?  From a transport theory point of view it indicates
that something is unphysical about the simulation since in transport theory the pressure components
are obtained from moments of the momentum-squared over the energy, e.g. for the longitudinal pressure
\beq
p_L = \int\!\frac{d^3p}{(2\pi)^3} \, \frac{p_z^2}{p^0} \, f({\bf p}) \, ,
\eeq
where $f({\bf p})$ is the one-particle phase-space distribution function.  Therefore, in transport theory
all components of the pressure are positive definite.
It is possible to generate negative longitudinal pressure in the case coherent fields as in the case of
the early-time evolution of the quark-gluon plasma \cite{Romatschke:2006wg,Fries:2007iy,%
Kovchegov:2007pq,Rebhan:2008uj};
however, such coherent fields are beyond the scope of hydrodynamical simulations which describe
the time evolution of a locally color- and charge-neutral fluid.

This fundamental issue aside, the negativity of the longitudinal pressure indicates that the expansion
which was used to derive the hydrodynamical equations themselves is breaking down.  This expansion
implicitly relies on the perturbation described by $\Pi$ being small compared to the isotropic pressure
$p$.  The point at which the longitudinal pressure goes to zero is the point at which the perturbation, 
$\Pi$, is equal in magnitude to the background around which one is expanding.  
This means that the perturbation is no longer a small correction to the system's evolution 
and that higher order corrections could become important. Therefore negative longitudinal pressure 
signals regions of parameter space where one cannot trust 2nd-order viscous hydrodynamical solutions.  
In the following two subsections we will make this statement quantitative and
extract constraints on the initial conditions which allow for 2nd-order viscous hydrodynamical
simulation.

\subsection{Determining the critical line in initial condition space}
\label{sec:criticalline}

For a fixed set of transport coefficients given by $\{\bar\eta, c_\pi, c_{\lambda_1}\}$ the only
remaining freedom in the hydrodynamical evolution equations (\ref{diffeqs}) comes from the coefficient
$\gamma$ (using the assumed ideal equation of state) and from the initial conditions through the 
dimensionless coefficient $k=\tau_0 \epsilon_0^{1/4}$ 
and the initial shear $\overline\Pi_0$.  In the next section we will vary these two parameters and
determine for which values one obtains a solution which, at any point during 
the evolution, has a negative longitudinal pressure.  For a given $\overline\Pi_0$ we find that for 
$k$ below a certain value, the system exhibits a negative longitudinal pressure.  We will define this
point in $k$ as the ``critical'' value of $k$.  Above the critical value of $k$ the longitudinal pressure
is positive definite at all times.

\subsubsection{Strong Coupling}
\label{sec:strongcriticalline}

\begin{figure*}[t]
\begin{center}
\includegraphics[width=10cm]{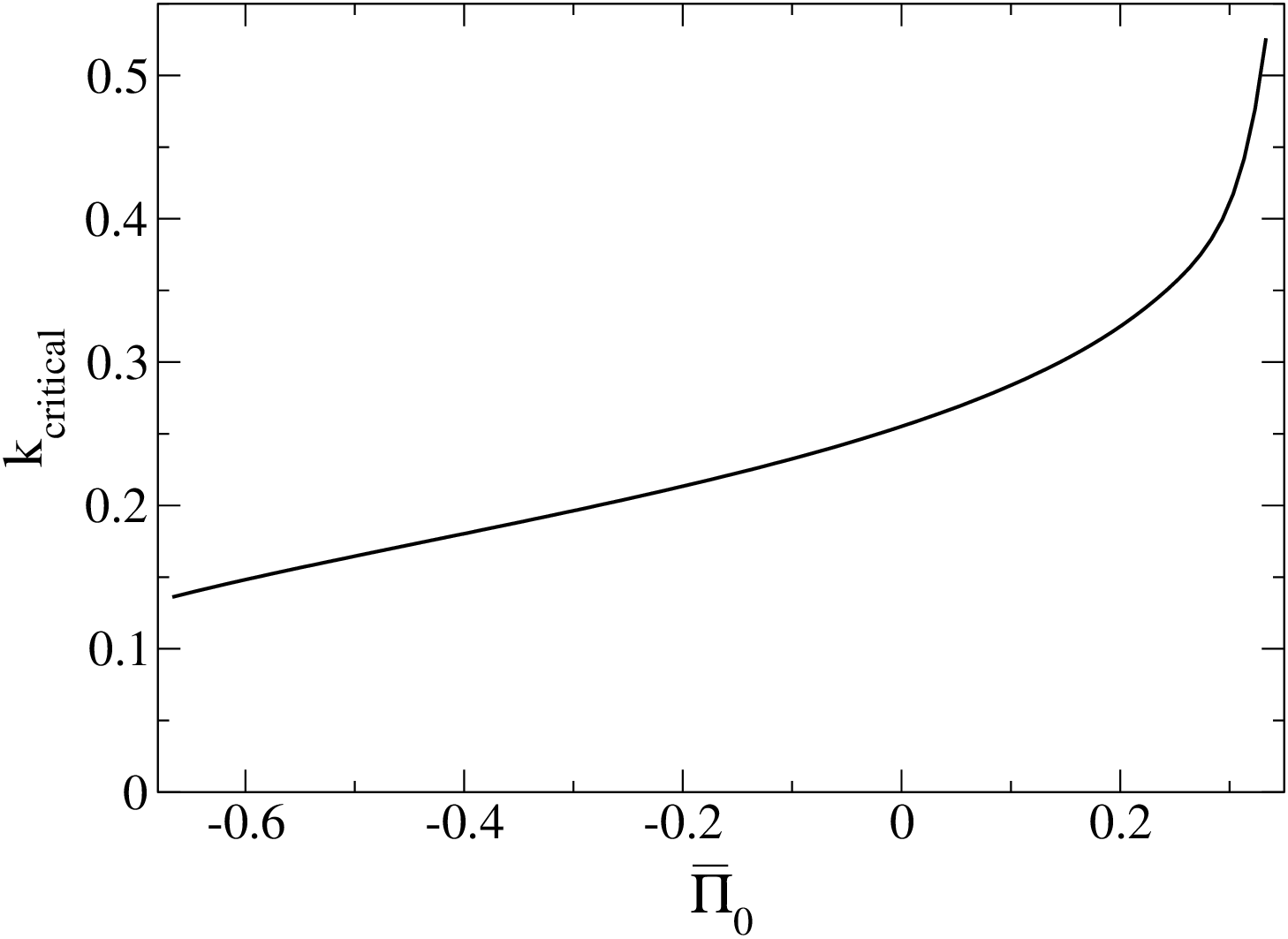}
\end{center}
\vspace{-6mm}
\caption{
Critical boundary in $k$ ($k_{\rm critical}$) as a function of the initial shear, $\overline\Pi_0$.  
Above this line solutions
have positive longitudinal pressure at all times.  Below this line solutions have negative longitudinal pressure at some
point during the evolution.  Transport coefficients in this case are the typical strong coupling 
values given in Eq.~(\ref{stronglimitvalues}). Left limit of plot region corresponds to $\Delta_0 = -1$ and right to 
$\Delta_0 = \infty$.
}
\label{fig:strongcouplingboundary}
\end{figure*}

In Fig.~\ref{fig:strongcouplingboundary} we plot the critical boundary in $k$ ($k_{\rm critical}$) as a function of the initial 
value of the shear, $\overline\Pi_0$.  Since $k$ is proportional to the assumed initial simulation time $\tau_0$
increasing $k$ with fixed initial energy density corresponds to increasing $\tau_0$.  Assuming fixed initial
temperature, for an initially prolate distribution, one can start the simulation at earlier times.  
For an initially oblate distribution, one must start the simulation at later times in order to remain above
the critical value of $k$.  In general, $k = \tau_0 \epsilon_0^{1/4}$ 
and our result can be used to set a bound on this product.

In the case of typical strong coupling transport coefficients, the critical value of $k$ at $\overline\Pi_0 = 0$ is 
$k_{\rm critical} (\overline\Pi_0 = 0) = 0.26$.  In the case of an ideal QCD equation of state and
assuming $\overline\Pi_0 = 0$, the constraint is that $\tau_0 > \gamma \, k_{\rm critical} \, T_0^{-1}$, which is numerically
$\tau_0 > 0.14 \, T_0^{-1}$.  Assuming an initial time of $\tau_0$ = 1 fm/c = 5.07 ${\rm GeV}^{-1}$ this implies that 
$T_0 > 28$ MeV. For other initial values of $\overline\Pi_0$ one can use Fig.~\ref{fig:strongcouplingboundary} to
determine the constraint.

\subsubsection{Weak Coupling}

\begin{figure*}[t]
\begin{center}
\includegraphics[width=10cm]{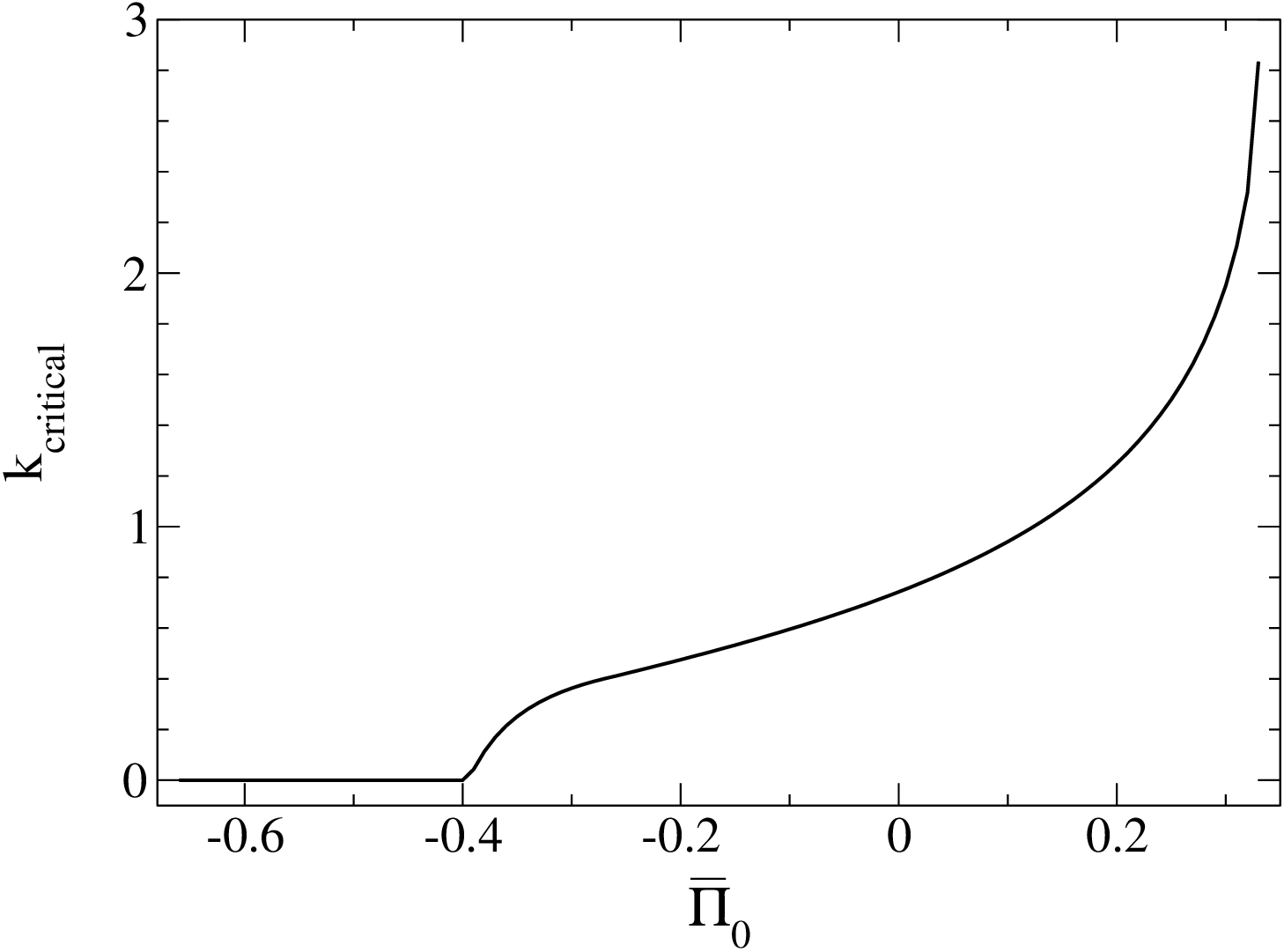}
\end{center}
\vspace{-6mm}
\caption{
Critical boundary in $k$ ($k_{\rm critical}$) as a function of the initial shear, $\overline\Pi_0$.  
Above this line solutions
have positive longitudinal pressure at all times.  Below this line solutions have negative longitudinal pressure at some
point during the evolution.  Transport coefficients in this case are the typical weak coupling 
values given in Eq.~(\ref{weaklimitvalues}). Left limit of plot region corresponds to $\Delta_0 = -1$ and right to 
$\Delta_0 = \infty$.
}
\label{fig:weakcouplingboundary}
\end{figure*}

In Fig.~\ref{fig:weakcouplingboundary} we plot the critical boundary in $k$ ($k_{\rm critical}$) as a function of the initial 
value of the shear, $\overline\Pi_0$.  Since $k$ is proportional to the assumed initial simulation time $\tau_0$
increasing $k$ with fixed initial energy density corresponds to increasing $\tau_0$.  As in the case of 
strong coupling, for an initially prolate distribution, one can start the simulation at earlier times.  For 
an initially oblate distribution, one must start the simulation at later times in order to remain above
the critical value of $k$.

In the case of typical weak coupling transport coefficients the critical value of $k$ at $\overline\Pi_0 = 0$ is $k_{\rm critical}
(\overline\Pi_0 = 0) = 0.74$.  In the case of an ideal QCD equation of state and
assuming $\overline\Pi_0 = 0$, the constraint is that
$\tau_0 > \gamma \, k_{\rm critical} \, T_0^{-1}$, which is numerically,
$\tau_0 > 0.40 \, T_0^{-1}$.  Assuming an initial time of $\tau_0$ = 1 fm/c this implies that 
$T_0 > 79$ MeV. For other initial values of $\overline\Pi_0$ one can use Fig.~\ref{fig:weakcouplingboundary} to
determine the constraint.

\subsection{For which initial conditions can one trust 2nd-order viscous hydrodynamical evolution?}
\label{sec:convergencecriteria}

As mentioned in Sec.~\ref{sec:momentumanisotropybounds} the requirement that the longitudinal 
pressure is positive during the simulated time only gives a weak constraint in the sense that it merely
requires that $\overline\Pi < \bar{p}$.  A stronger constraint can be obtained by requiring instead
$- \alpha \, \bar{p} \leq \overline\Pi \leq \alpha \, \bar{p}$ and then using this to constrain the
possible initial time and energy density which can be used in hydrodynamical simulations.  In the
following subsections we will fix $\alpha=1/3$ as our definition of what is a ``large'' correction.
For this value of $\alpha$ the initial values of $\overline\Pi_0$ are constrained to be between
$-1/9 \leq \overline\Pi_0 \leq 1/9$.  For a given $\overline\Pi_0$ in this range we find that for 
$k$ below a certain value we cannot satisfy the stronger constraint at all simulated times.  We will 
define this point in $k$ as the ``convergence'' value of $k$ or $k_\text{convergence}$.  Above this 
value of $k=k_\text{convergence}$ the shear satisfies the constraint 
$- \bar{p}/3 \leq \overline\Pi \leq \bar{p}/3$ at all simulated times and therefore represents a
``reasonable'' simulation.

\subsubsection{Strong Coupling}

\begin{figure*}[t]
\begin{center}
\includegraphics[width=10cm]{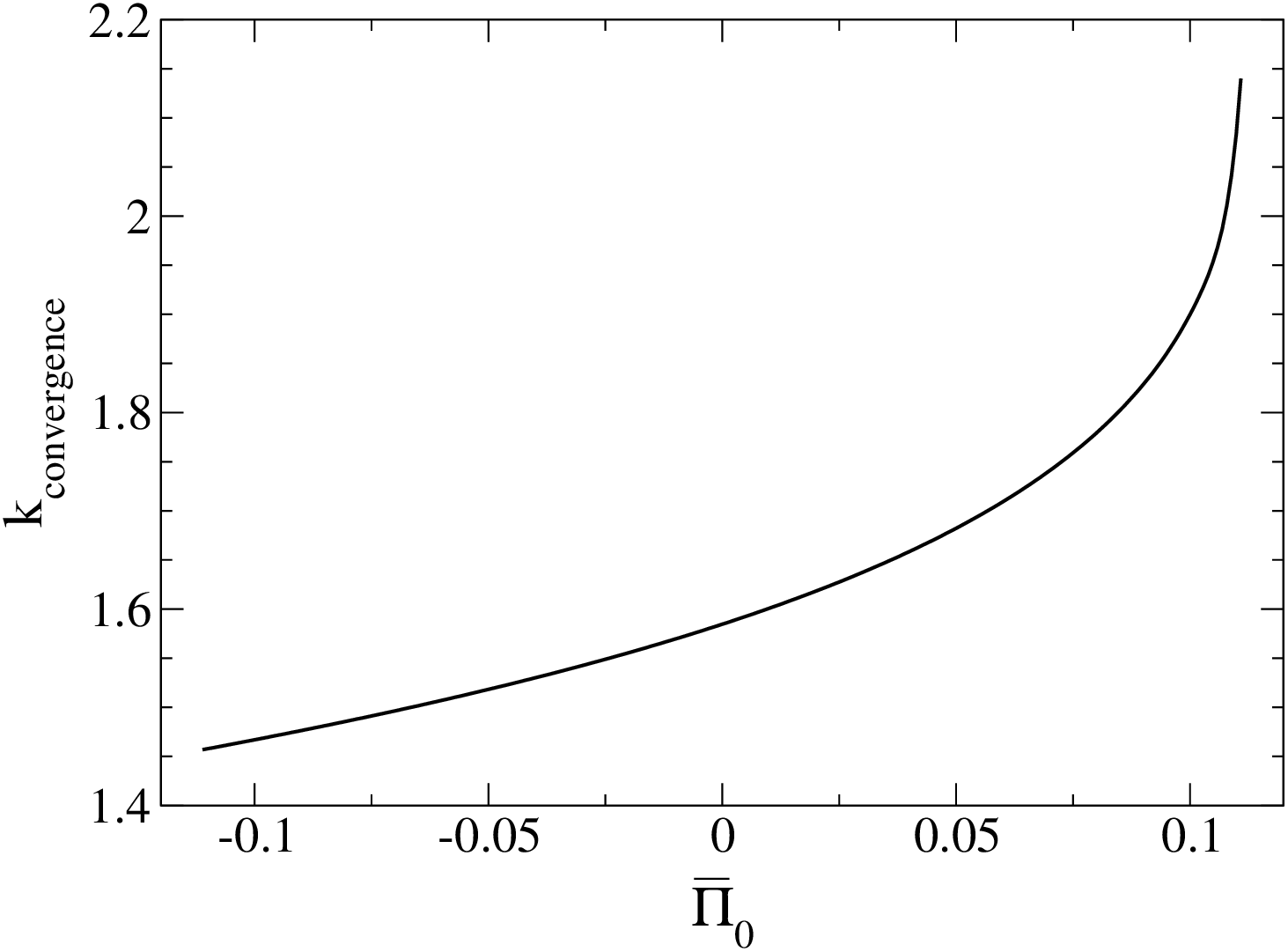}
\end{center}
\vspace{-6mm}
\caption{
Convergence boundary in $k$ ($k_{\rm convergence}$) as a function of the initial shear, $\overline\Pi_0$.  
Above this line solutions
satisfy the convergence constraint.  Transport coefficients in this case are the typical strong coupling 
values given in Eq.~(\ref{stronglimitvalues}).}
\label{fig:strongcouplingconvergence}
\end{figure*}

In Fig.~\ref{fig:strongcouplingconvergence} we plot the ``convergence boundary'' in $k$ ($k_{\rm convergence}$) 
as a function of the initial shear, $\overline\Pi_0$.
In the case of typical strong coupling transport coefficients the convergence value of $k$ at $\overline\Pi_0 = 0$ is $k_{\rm 
convergence}(\overline\Pi_0 = 0) = 1.58$.  In the case of an ideal 
QCD equation of state and assuming $\overline\Pi_0 = 0$, the constraint is that 
$\tau_0 > \gamma \, k_{\rm convergence} \, T_0^{-1}$, which is numerically $\tau_0 > 0.85 \, T_0^{-1}$.  
Assuming an initial time of $\tau_0$ = 1 fm/c  this implies that $T_0 > 167$ MeV. For other initial values of 
$\overline\Pi_0$ one can use Fig.~\ref{fig:strongcouplingconvergence} to determine the constraint.

\subsubsection{Weak Coupling}

\begin{figure*}[t]
\begin{center}
\includegraphics[width=10cm]{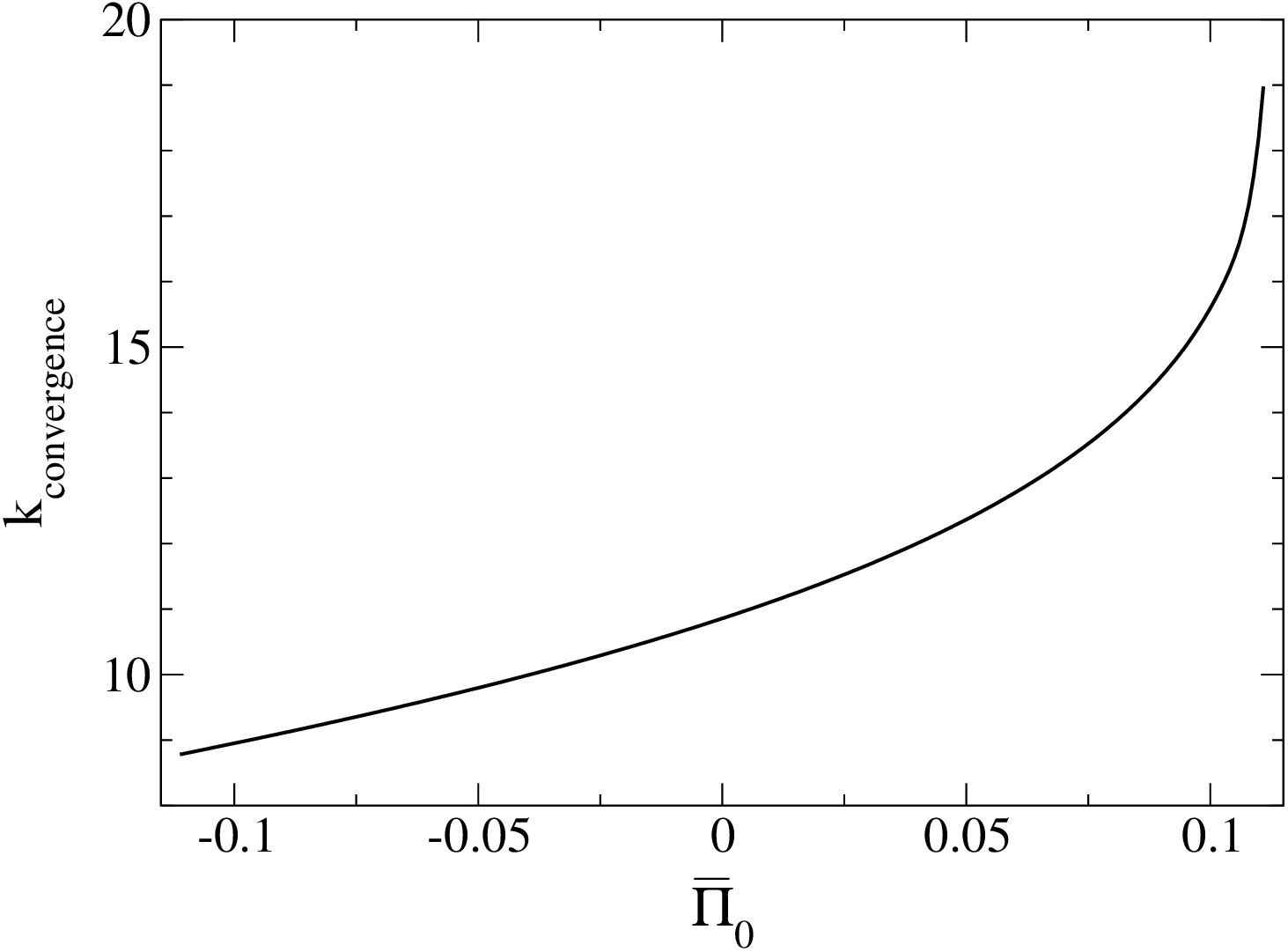}
\end{center}
\vspace{-6mm}
\caption{
Convergence boundary in $k$ ($k_{\rm convergence}$) as a function of the initial shear, $\overline\Pi_0$.  
Above this line solutions
satisfy the convergence constraint.  Transport coefficients in this case are the  typical weak coupling 
values given in Eq.~(\ref{weaklimitvalues}).}
\label{fig:weakcouplingconvergence}
\end{figure*}

In Fig.~\ref{fig:weakcouplingconvergence} we plot the ``convergence boundary'' in $k$ ($k_{\rm convergence}$) 
as a function of the initial shear, $\overline\Pi_0$.
In the case of typical weak coupling transport coefficients the convergence value of $k$ at $\overline\Pi_0 = 0$ is $k_{\rm 
convergence}(\overline\Pi_0 = 0) = 10.9$. In the case of an ideal QCD equation of state and
assuming $\overline\Pi_0 = 0$, the constraint is that
$\tau_0 > \gamma \, k_{\rm convergence} \, T_0^{-1}$, which is numerically 
$\tau_0 >  5.9 \, T_0^{-1}$.    Assuming an initial time of $\tau_0$ = 1 fm/c = 5.07 ${\rm GeV}^{-1}$ 
this implies that $T_0 > 1.16$ GeV.  For other initial values of $\overline\Pi_0$ one can use 
Fig.~\ref{fig:weakcouplingconvergence} to determine the constraint.

\subsection{What does this imply for higher dimensional hydrodynamical simulations?}
\label{sec:higherd}

If one proceeds to more realistic simulations in higher dimensional boost invariant treatments, e.g. 1+1 and 2+1,
the spatial variation of the initial conditions and time evolution in the transverse
plane have to be taken into account.  In addition, new freedoms
such as the initial fluid flow field and additional transport coefficients arise; however,
to first approximation one can treat these higher dimensional systems as a collection of 0+1 
dimensional systems with different initial conditions at each point in the transverse plane.  
Within this approximation one would quickly
find that there are problems with the hydrodynamic treatment at the transverse edges of the simulated
region.  

This happens because as one goes away from the center of the hot and dense matter, the energy density
(temperature) drops and, assuming a fixed initial simulation time $\tau_0$, one would find that
at a finite distance from the center the condition $k > k_\text{critical}$ would be violated by the initial
conditions.  In these
regions of space, hydrodynamics would then predict an infinitely large anisotropy parameter, $\Delta$,
casting doubt on the reliability of the hydrodynamic assumptions.  Even worse is that at a smaller distance 
from the center one would cross the ``convergence boundary'' in $k$, $k_\text{convergence}$, and therefore 
not fully trust the analytic approximations used in deriving the hydrodynamic equations 
(conformality, truncation at 2nd order, etc.).

Of course, an approximation by uncoupled 0+1 systems with different initial conditions 
would not generate any radial or elliptic flow;
however, we find empirically that the picture above holds true in higher-dimensional simulations,
justifying the basic logic.  For example, using strongly-coupled transport coefficients and assuming
an initially isotropic plasma ($\overline\Pi_0=0$), we found in Sec.~\ref{sec:strongcriticalline} that 
$k_{\rm critical} = 0.26$.  
In terms of the initial temperature this predicts that when starting a simulation with $\tau_0=1$ fm/c, one
will generate negative longitudinal pressures for any initial temperature $T_0 \lsim 28$ MeV.

\begin{figure*}[t]
\begin{center}
\includegraphics[width=8cm]{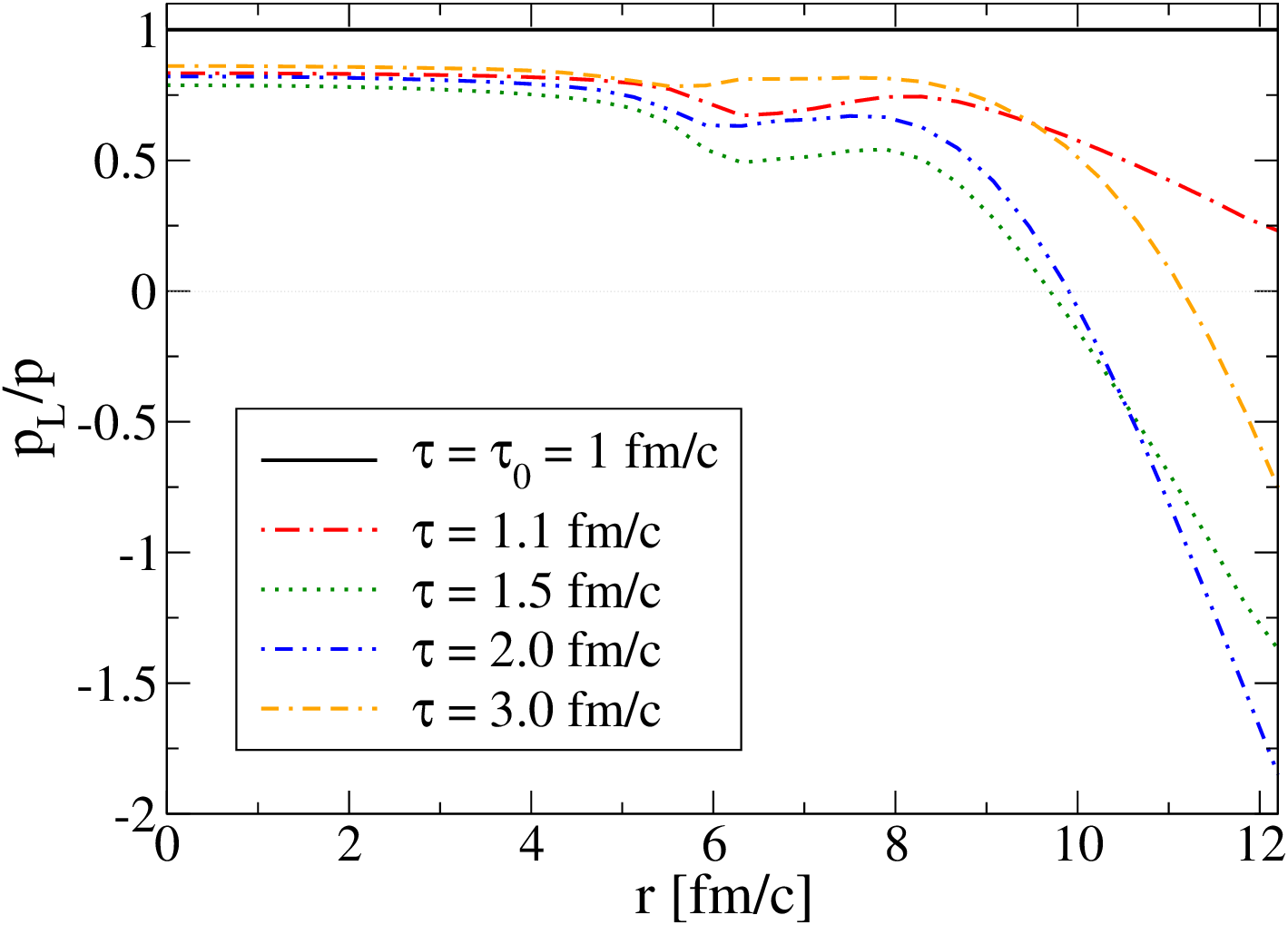}
$\;\;$
\includegraphics[width=8cm]{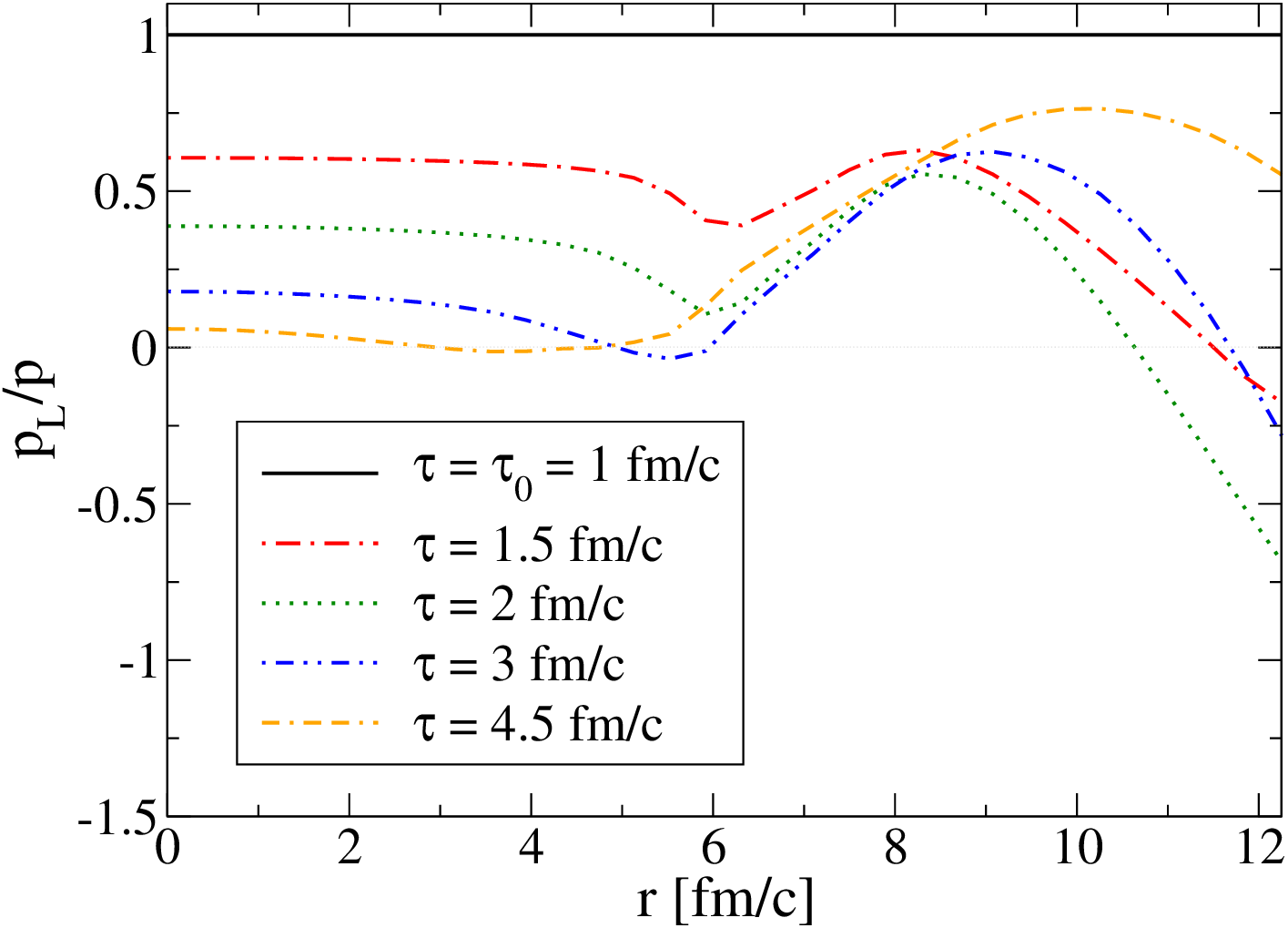}
\end{center}
\vspace{-6mm}
\caption{
Evolution of the longitudinal pressure in proper-time obtained from the 2+1 dimensional viscous hydrodynamics
code of Ref.~\cite{Luzum:2008cw}.  Horizontal axis is the distance from the center of the simulated
region.  In the left panel we show the result obtained using the typical strong coupling values given in 
Eq.~(\ref{stronglimitvalues}) but with $c_{\lambda_1}=0$.  In the right panel we show the result obtained using 
the typical weak coupling values given in 
Eq.~(\ref{weaklimitvalues}) but with $c_{\lambda_1}=0$.  The runs shown used Glauber initial conditions with
an initial central temperature of $T_0 = 350$ MeV, initial time $\tau_0 = 1$ fm/c 
and $\Pi_\mu^\nu(\tau_0)=0$.
}
\label{fig:hydro1p1}
\end{figure*}

We will now compare this prediction with results for the longitudinal pressure extracted from the 2+1 
dimensional code of Luzum and Romatschke \cite{Luzum:2008cw,RomatschkeCode}.  
In Fig.~\ref{fig:hydro1p1} we show fixed $\tau$
snapshots of the longitudinal pressure.
The runs shown in Fig.~\ref{fig:hydro1p1} were 
performed on a $69^2$ transverse lattice with a lattice spacing of 2 ${\rm GeV}^{-1}$ using 
Glauber initial conditions starting at $\tau_0$=1 fm/c, an initial central temperature of 
$T_0 = 350$ MeV, zero initial shear and zero impact parameter.  For these runs we have
used the realistic QCD equation of state used in Ref.~\cite{Luzum:2008cw}.
In the left panel of Fig.~\ref{fig:hydro1p1} the transport coefficients 
were set to the typical strong coupling values given in Eq.~(\ref{stronglimitvalues}), except with 
$c_{\lambda_1}=0$ due to the fact that the code used did not include this term in the hydrodynamic 
equations.
Based on the initial transverse temperature profile and our estimated critical initial temperature, 
in the strong-coupling case we expect 
negative longitudinal pressures to be generated at transverse radius $r \gsim 10$ fm.
As can be seen from the left panel of Fig.~\ref{fig:hydro1p1}, at the edge of the simulated region 
the longitudinal pressure becomes negative starting already at very early times.  The transverse radii at 
which this occurs is in good agreement with our estimate based on the 0+1 dimensional critical 
value detailed above.

Based on our convergence criterium detailed in
Sec.~\ref{sec:convergencecriteria} we found, in the strong-coupling case, that $k_{\rm 
convergence}(\overline\Pi_0 = 0) = 1.58$.  Assuming $\tau_0 = 1$ fm/c this translates into a
minimum initial temperature of 167 MeV.  Based on the transverse temperature profile 
used in the run shown in the left panel of Fig.~\ref{fig:hydro1p1} this results in
a maximum transverse radius $r \sim 6.8$ fm.  At radii larger than this value it
is possible that higher order corrections are large and therefore the applicability of 2nd-order viscous
hydrodynamics becomes questionable.  Since this temperature is greater than the typical
freeze-out temperature used, $T_f \sim 150$ MeV, this means that in the strong coupling
limit it is relatively safe to use hydrodynamical simulations.  However, one should be
extremely careful with the transverse edges.

The situation, however, is not as promising in the weak-coupling case.  To see
this explicitly, in the right panel of Fig.~\ref{fig:hydro1p1} we show the longitudinal 
pressure resulting from a run with weak
coupling transport coefficients (\ref{weaklimitvalues}).  
Based on the initial transverse temperature profile and our estimated critical initial temperature, 
in the weak-coupling case we expect 
negative longitudinal pressures to be generated at transverse radius $r \gsim 8$ fm.
Comparing this prediction to the results shown in the right panel of Fig.~\ref{fig:hydro1p1}
we see that the situation is even worse than expected.  By the final time of 4.5 fm/c the entire
central region has very low or negative longitudinal pressure.  We note that at that time the radius at
which the temperature has dropped below the freeze-out temperature is around 7.3 fm so
the region where the longitudinal pressure is negative (or almost negative) is still in the QGP phase.

In terms of convergence, we remind the reader that based on our convergence 
criterium detailed in Sec.~\ref{sec:convergencecriteria} we found 
that in the weakly-coupled case 
$k_{\rm convergence}(\overline\Pi_0 = 0) = 10.9$.  Assuming $\tau_0 = 1$ fm/c we found 
that the initial central temperature should be greater than 1.16 GeV.  As can be seen
in Fig.~\ref{fig:hydro1p1} the corrections to ideal hydrodynamics are sizable so this 
again points to the possibility that there are large corrections to the 2nd-order hydrodynamic
equations.  Based
on this, it would be questionable to ever apply 2nd-order viscous hydrodynamics
to a weakly-coupled quark-gluon plasma generated in relativistic heavy-ion collisions.
At the very least one would need to include
nonconformal 2nd-order terms and 3rd-order terms in order to assess their impact. 

\section{Conclusions and Outlook}
\label{sec:conclusions}

In this paper we have derived two general criteria that can be used to assess the applicability
of 2nd-order conformal viscous hydrodynamics to relativistic heavy-ion collisions.  We did this
by simplifying to a 0+1 dimensional system undergoing boost invariant expansion 
and then (a) requiring the longitudinal pressure to
be positive during the simulated time or (b) requiring a convergence criterium that $|\Pi| < p/3$ 
during the simulated time.  We showed that these requirements lead to a non-trivial relation between 
the possible initial simulation time $\tau_0$, the initial energy density $\epsilon_0$, and the initial 
value of the fluid shear tensor, $\Pi_0$.  As a cross check of our numerics we presented an approximate 
analytic solution of 2nd-order conformal viscous hydrodynamical evolution which represents the leading
correction to 0+1 dimensional boost-invariant ideal hydrodynamics in the limit $\eta/s \rightarrow 0$.

The constraints derived here were then shown to provide guidance for where one might expect
2nd-order viscous hydrodynamics to be a good approximation in higher-dimensional cases.  
We found that the prediction
of our criticality bound was in reasonable agreement with where the longitudinal 
pressure becomes negative in 2+1 dimensional viscous hydrodynamical simulations.  Based on these
findings it seems possible to estimate where one obtains convergent/trustable 2nd-order viscous
hydrodynamical simulations based solely on the initial conditions and analysis of the hydrodynamical
evolution equations themselves.

In closing we mention that another outcome of this work is that 
we have shown that it is possible to use hydrodynamical simulations
to predict the proper-time dependence of the plasma momentum-space
anisotropy as quantified by the $\Delta$ or $\xi$ parameters.
This can be used as input to calculations of production of electromagnetic radiation from an
anisotropic plasma \cite{Mauricio:2007vz, Martinez:2008di, Martinez:2008rr, Martinez:2008mc,%
Schenke:2006yp, Bhattacharya:2008up, Bhattacharya:2008mv}, calculations of 
quarkonium binding/polarization in anisotropic plasma \cite{Dumitru:2007hy,Dumitru:2009ni}, 
and also to assess the phenomenological growth rate of plasma instabilities on top of the mean 
colorless fluid background (see Ref.~\cite{Strickland:2007fm} and references therein).
The findings here present a complication in this regard since phenomenological studies
will require knowledge of $\Delta$ in the full transverse plane.  As we have shown, 
2nd-order hydrodynamical simulations predict that this parameter can become
infinite in certain regions.  In these regions one would no longer trust the predictions
of the hydrodynamical model and additional input would be required. 

\section*{Acknowledgements}

We are extremely grateful to Pasi Huovinen for his careful reading of our manuscript.
We also acknowledge conversations with Adrian Dumitru, Berndt M\"uller, 
Paul Romatschke, and Derek Teaney.  M. Martinez thanks J. Aichelin and K. Eskola 
for assistance provided in order to attend the 18th Jyvaskyla Summer School 2008 where this work 
was initiated. M. Martinez was supported by the Helmholtz Research School and Otto Stern School 
of the Goethe-Universit\"at Frankfurt am Main. M. Strickland was supported partly by the Helmholtz International
Center for FAIR Landesoffensive zur Entwicklung Wissenschaftlich-\"Okonomischer Exzellenz program.


\appendix

\section{Notation and conventions}
\label{app:notations}
We summarize the conventions and notation we use in the main body of the text:
\begin{itemize}
 \item The metric for a Minkowski space in the curvilinear coordinates $(\tau, x,y,\zeta)$ is $g_{\mu \nu}={\rm diag}(g_{\tau \tau},g_{xx},g_{yy},g_{\zeta \zeta})=(1,-1,-1,-\tau^2)$ \, .
 \item $\Delta^{\mu \nu}=g^{\mu \nu}-u^{\mu}u^\nu$ is a projector orthogonal to the fluid velocity, 
$u_\mu\Delta^{\mu \nu}=0$.
 \item The comoving time derivative: $D\equiv u^\alpha D_\alpha$.
 \item The comoving space derivative: $\nabla^\mu\equiv \Delta^{\mu \alpha} D_\alpha$.
 \item The brackets $\langle\ \rangle$ denote an operator that is symmetric, traceless, and orthogonal to the fluid velocity:
\begin{equation}
A_{\langle \mu} B_{\nu\rangle} =\left(\Delta^\alpha_\mu \Delta^\beta_\nu + 
\Delta^\alpha_\nu \Delta^\beta_\mu-\frac{2}{3} \Delta^{\alpha \beta} 
\Delta_{\mu \nu}\right) A_\alpha B_\beta \, .
\end{equation}
 \item The symmetric and anti-symmetric operators:
\begin{eqnarray}
A_{(\mu} B_{\nu)} &=&\frac{1}{2}\left(A_\mu B_\nu+A_\nu B_\mu\right) \, ,
\\
A_{[\mu} B_{\nu]} &=&\frac{1}{2}\left(A_\mu B_\nu-A_\nu B_\mu\right) \, .
\end{eqnarray}
\end{itemize}

\section{Relation between $\Delta$ and $\xi$}
\label{app:xideltarelation}

In this appendix we derive the relation between the anisotropy parameter $\Delta$ introduced
in this paper and the $\xi$ parameter introduced in Ref.~\cite{Romatschke:2003ms}.  In the 
general case $\xi$ is defined by taking an arbitrary isotropic distribution function $f_\text{iso}(p)$
and stretching or squeezing it along one direction in momentum space to obtain an anisotropic distribution.
Mathematically this is done by introducing a unit vector ${\bf\hat{n}}$ which defines the direction of
anisotropy, an anisotropy parameter $-1 < \xi < \infty$, and requiring $f({\bf p}) = f_\text{iso}\!\left(\sqrt{{\bf p}^2 +
\xi ({\bf p}\cdot{\bf\hat{n}})^2}\right)$.  Fixing ${\bf\hat{n}} = {\bf\hat{z}}$ to define the longitudinal
direction and assuming massless particles, is it straightforward to evaluate the transverse and longitudinal 
pressures through the components of the stress-energy tensor
\beq
p_T = \frac{1}{2}\left( T^{xx} + T^{yy}\right) = \frac{1}{2} \int \frac{d^3{\bf p}}{(2\pi)^3} \, 
\frac{p_x^2 + p_y^2}{|{\bf p}|} \,
f_\text{iso}\!\left(\sqrt{{\bf p}^2 +\xi p_z^2}\right) \, ,
\eeq
and
\beq
p_L = T^{zz} = \int \frac{d^3{\bf p}}{(2\pi)^3} \, 
\frac{p_z^2}{|{\bf p}|} \,
f_\text{iso}\!\left(\sqrt{{\bf p}^2 +\xi p_z^2}\right) \, .
\eeq
By a change of variables to $\tilde{p} \equiv \sqrt{{\bf p}^2 +\xi p_z^2}$
and the use of spherical coordinates one can show that
\beq
p_T = \frac{3}{4 \xi} \left( 1 + (\xi-1)\frac{\text{atan}\sqrt{\xi}}{\sqrt{\xi}} \right) p_T^\text{iso} \, ,
\eeq
and
\beq
p_L = \frac{3}{2 \xi} \left( \frac{\text{atan}\sqrt{\xi}}{\sqrt{\xi}} - \frac{1}{1+\xi} \right) p_L^\text{iso} \, ,
\eeq
where $p_T^\text{iso}$ and $p_L^\text{iso}$ are the isotropic transverse and longitudinal pressures
which are obtained from $f_\text{iso}$, respectively.
Combining the above relations and using $p_T^\text{iso} = p_L^\text{iso} = \epsilon^\text{iso}/3$, 
where $\epsilon^\text{iso}$ is the isotropic energy density, we obtain the following expression for 
$\Delta$
\beq
\Delta = \frac{1}{2} \left(\xi-3\right) + \xi \left(  (1+\xi)\frac{\text{atan}\sqrt{\xi}}{\sqrt{\xi}} - 1 \right)^{-1}\, .
\label{deltaxi}
\eeq 
In the small $\xi$ limit
\beq
\lim_{\xi \rightarrow 0} \Delta = \frac{4}{5} \xi + {\cal O}(\xi^2) \, ,
\eeq
and in the large $\xi$ limit
\beq
\lim_{\xi \rightarrow \infty} \Delta = \frac{1}{2} \xi + {\cal O}(\sqrt{\xi}) \, .
\eeq
For general $\xi$ one needs to invert (\ref{deltaxi}) numerically in order to obtain
$\xi$ as a function of $\Delta$.

\bibliography{hydrocriticality}

\end{document}